\documentclass[11pt]{article}
\pdfoutput=1
\usepackage{amsfonts,amsmath}
\usepackage{amssymb}
\usepackage{fancyhdr}
\usepackage{slashed}
\usepackage{graphicx}
\usepackage{subfigure}
\usepackage{color}

\usepackage{hyperref}
\usepackage[utf8]{inputenc}
\usepackage[titletoc]{appendix}

\def\hybrid{\topmargin -20pt    \oddsidemargin 0pt
        \headheight 0pt \headsep 0pt
        \textwidth 6.25in       
        \textheight 9.25in       
        \marginparwidth .875in
        \parskip 5pt plus 1pt   \jot = 1.5ex}

\hybrid

\def\baselinestretch{1.2}

\catcode`\@=11

\def\marginnote#1{}
\newcount\hour
\newcount\minute
\newtoks\amorpm
\hour=\time\divide\hour by60
\minute=\time{\multiply\hour by60 \global\advance\minute by-\hour}
\edef\standardtime{{\ifnum\hour<12 \global\amorpm={am}%
        \else\global\amorpm={pm}\advance\hour by-12 \fi
        \ifnum\hour=0 \hour=12 \fi
        \number\hour:\ifnum\minute<10 0\fi\number\minute\the\amorpm}}
\edef\militarytime{\number\hour:\ifnum\minute<10 0\fi\number\minute}

\def\draftlabel#1{{\@bsphack\if@filesw {\let\thepage\relax
   \xdef\@gtempa{\write\@auxout{\string
      \newlabel{#1}{{\@currentlabel}{\thepage}}}}}\@gtempa
   \if@nobreak \ifvmode\nobreak\fi\fi\fi\@esphack}
        \gdef\@eqnlabel{#1}}
\def\@eqnlabel{}
\def\@vacuum{}
\def\draftmarginnote#1{\marginpar{\raggedright\scriptsize\tt#1}}

\def\draft{\oddsidemargin -.5truein
        \def\@oddfoot{\sl preliminary draft \hfil
        \rm\thepage\hfil\sl\today\quad\militarytime}
        \let\@evenfoot\@oddfoot \overfullrule 3pt
        \let\label=\draftlabel
        \let\marginnote=\draftmarginnote
   \def\@eqnnum{(\theequation)\rlap{\kern\marginparsep\tt\@eqnlabel}%
\global\let\@eqnlabel\@vacuum}  }

\def\preprint{\twocolumn\sloppy\flushbottom\parindent 2em
        \leftmargini 2em\leftmarginv .5em\leftmarginvi .5em
        \oddsidemargin -.5in    \evensidemargin -.5in
        \columnsep .4in \footheight 0pt
        \textwidth 10.in        \topmargin  -.4in
        \headheight 12pt \topskip .4in
        \textheight 6.9in \footskip 0pt
        \def\@oddhead{\thepage\hfil\addtocounter{page}{1}\thepage}
        \let\@evenhead\@oddhead \def\@oddfoot{} \def\@evenfoot{} }

\def\numberbysection{\@addtoreset{equation}{section}
        \def\theequation{\thesection.\arabic{equation}}}

\def\underline#1{\relax\ifmmode\@@underline#1\else
        $\@@underline{\hbox{#1}}$\relax\fi}

\def\titlepage{\@restonecolfalse\if@twocolumn\@restonecoltrue\onecolumn
     \else \newpage \fi \thispagestyle{empty}\c@page\z@
        \def\thefootnote{\fnsymbol{footnote}} }

\def\endtitlepage{\if@restonecol\twocolumn \else \newpage \fi
        \def\thefootnote{\arabic{footnote}}
        \setcounter{footnote}{0}}  

\catcode`@=12
\relax

\def\figcap{\section*{Figure Captions\markboth
        {FIGURECAPTIONS}{FIGURECAPTIONS}}\list
        {Figure \arabic{enumi}:\hfill}{\settowidth\labelwidth{Figure
999:}
        \leftmargin\labelwidth
        \advance\leftmargin\labelsep\usecounter{enumi}}}
 \relax
\def\tablecap{\section*{Table Captions\markboth
        {TABLECAPTIONS}{TABLECAPTIONS}}\list
        {Table \arabic{enumi}:\hfill}{\settowidth\labelwidth{Table
999:}
        \leftmargin\labelwidth
        \advance\leftmargin\labelsep\usecounter{enumi}}}
 \relax
\def\reflist{\section*{References\markboth
        {REFLIST}{REFLIST}}\list
        {[\arabic{enumi}]\hfill}{\settowidth\labelwidth{[999]}
        \leftmargin\labelwidth
        \advance\leftmargin\labelsep\usecounter{enumi}}}
 \relax
\makeatletter
\newcounter{pubctr}
\def\publist{\@ifnextchar[{\@publist}{\@@publist}}
\def\@publist[#1]{\list
        {[\arabic{pubctr}]\hfill}{\settowidth\labelwidth{[999]}
        \leftmargin\labelwidth
        \advance\leftmargin\labelsep
        \@nmbrlisttrue\def\@listctr{pubctr}
        \setcounter{pubctr}{#1}\addtocounter{pubctr}{-1}}}
\def\@@publist{\list
        {[\arabic{pubctr}]\hfill}{\settowidth\labelwidth{[999]}
        \leftmargin\labelwidth
        \advance\leftmargin\labelsep
        \@nmbrlisttrue\def\@listctr{pubctr}}}
 \relax
\makeatother
\newskip\humongous \humongous=0pt plus 1000pt minus 1000pt

\newif\ifdtup

\relax

\def\be{\begin{equation}}
\def\ee{\end{equation}}
\def\ba{\begin{eqnarray}}
\def\ea{\end{eqnarray}}

\def\k{\kappa}

\def\a{\alpha}

\def\b{\beta}

\def\g{\gamma}
\def\G{\Gamma}
\def\d{\delta}
\def\D{\Delta}
\def\e{\epsilon}

\def\th{\theta}

\def\m{\mu}
\def\n{\nu}
\def\om{\omega}

\def\s{\sigma}
\def\S{\Sigma}

\def\cN{{\cal N}}

\def\trr{{\tilde{r}}}
\def\tom{{\tilde{\omega}}}

 \def\cN{{\cal N}} \def\cO{{\cal O}}

\newcommand{\vev}[1]{{\left< {#1} \right>}}

\newcommand{\ket}[1]{{\left| {#1} \right>}}
\newcommand{\prt}[1]{{\left( {#1} \right)}}
\newcommand{\prtt}[1]{{\left[ {#1} \right]}}

\def\no{\noindent}

\def\IR{\relax{\rm I\kern-.18em R}}

\def\pp{\partial}
\newcommand{\ti}{\tilde}
\newcommand{\ff}{\frac}

\def \ta  { {\tilde {a }}}

\def\IR{\relax{\rm I\kern-.18em R}}
\def\IL{\relax{\rm I\kern-.18em L}}

\def\inv{^{\raise.15ex\hbox{${\scriptscriptstyle -}$}\kern-.05em 1}}

\def\bea{\begin{eqnarray}}
\def\eea{\end{eqnarray}}
\newcommand{\eq}[1]{(\ref{#1})}
\def\nn{\nonumber}

\newcommand{\la}[1]{\label{#1}}

\def\a{\alpha}      
\def\b{\beta}       
\def\g{\gamma}  \def\G{\Gamma}  
\def\d{\delta}  \def\D{\Delta}  
\def\e{\epsilon}

\def\k{\kappa}
 
\def\m{\mu} \def\n{\nu}
\def\o{\omega}

\def\s{\sigma}  \def\S{\Sigma}
\def\t{\tau}
\def\th{\theta}

\def \dx {\dot{x}}

\definecolor{markcolor2}{rgb}{1,0,0}

\definecolor{markcolor3}{rgb}{0,1,0}

\newcommand{\rd}{\partial}

\begin{document}

\renewcommand{\theequation}{\thesection.\arabic{equation}}
\csname @addtoreset\endcsname{equation}{section}

\newcommand{\beq}{\begin{equation}}
\newcommand{\eeq}[1]{\label{#1}\end{equation}}
\newcommand{\ber}{\begin{eqnarray}}
\newcommand{\eer}[1]{\label{#1}\end{eqnarray}}
\newcommand{\eqn}[1]{(\ref{#1})}
\begin{titlepage}

\begin{center}

~
\vskip 1 cm

{\Large
\bf  Quantum Fluctuation and Dissipation in Holographic Theories: A Unifying Study Scheme
}

\vskip 0.5in

 {\bf Dimitrios Giataganas$^{1}$~, \phantom{x}  Da-Shin Lee$^{2}$~,  \phantom{x}  Chen-Pin Yeh}$^{2 }$
 \vskip 0.1in
 {\em
   ${}^1$  Physics Division, National Center for Theoretical
   Sciences, \\
  National Tsing-Hua University, Hsinchu, 30013, Taiwan
 \vskip .15in
 ${}^2$
 Department of Physics, National Dong-Hwa University, Hualien 97401, Taiwan, R.O.C.
 \\\vskip .1in
 {\tt ~  dimitrios.giataganas@cts.nthu.edu.tw, dslee@gms.ndhu.edu.tw, chenpinyeh@gms.ndhu.edu.tw
 }\\
 }

\vskip .2in
\end{center}

\vskip .4in

\centerline{\bf Abstract}
Motivated by the wide range of applicability of the fluctuation and dissipation phenomena in non-equilibrium systems, we provide a universal study scheme for the dissipation of the energy and the corresponding Brownian motion analysis of massive particles due to quantum and thermal fluctuations in a wide class of strongly coupled quantum field theories. The underlying reason for the existence of such  unified study scheme, is that our analytic methods turn out to heavily depend on the order of the Bessel functions $\nu$, describing the string fluctuations attached to the particle. Different values of the order are associated to different theories.

The two-point function of the fluctuations exhibits two different late time behaviors, depending purely on the value of the order of Bessel functions. We then find that the coefficients and observables associated with the stochastic motion at zero and finite temperature, depend on the scales of the theory through powers of the order $\nu$.  Moreover, the fluctuation-dissipation theorem is verified from the bulk perspective to be universally satisfied for the whole class of theories.
Finally, we show that the analysis of certain types of Dp-brane fluctuations can be mapped one-to-one to the string fluctuations and therefore the stochastic brane observables can be read from the string ones. In the closing remarks we demonstrate how our analysis accommodates known results as special cases and provide more applications.

\no
\end{titlepage}
\vfill
\eject


\noindent


\def\baselinestretch{1.2}
\baselineskip 19 pt
\noindent


\setcounter{equation}{0}

\section{Introduction}

\textbf{The Wide Range of Applicability of Non-Equilibrium Theories of Brownian motion:}
Irreversibility in nonequilibrium processes is an ubiquitous phenomenon.  The exploration of the effect  of dissipation and the associated dynamics of a non-equilibrium system that evolves toward thermal equilibrium from the microscopic point of view is of fundamental importance for understanding the origin of irreversibility.   The Brownian motion, historically observed due the irregular dynamics, is exhibited by a test particle in a liquid environment, which is caused by random microscopic interactions/kicks with the particles of the liquid. The initial theoretical developments explaining the Brownian motion served at the time  as the final proof of the existence of atoms and provided a solid way to determine the value of the Avogadro constant. At the same time, the probabilistic description was introduced, laying down the foundation of nonequilibrium statistical mechanics with fundamental equations  such as the Langevin, and the Fokker-Planck equations to name but a few. Since then the study  on nonequilibrium phenomena
remains very active due to its ubiquitous feature  on a variety of systems and applications \cite{brownianII,brownian100,brownianrel,brownian111}.

The key idea behind the physical behavior of such out of equilibrium phenomena and their tendency toward equilibrium, is the focus on their statistical description and secondary on the nature of interactions between the system and environment.  The environment has certain characteristics such as an infinite specific heat, a condition necessary for the equivalence between a microcanonical and canonical description, and thus infinitely many degrees of freedom. Thus, when the system initially is away from equilibrium with the environment,  according to statistical mechanics the effects from the environment to the time-evolved system can be described in terms of equilibrium expectation values or correlators of the environmental degrees of freedom at finite or zero temperature. In particular, the linear response theory gives a prime example that by displacing slightly away of equilibrium a macroscopic variable, the system's  restoring force is proportional linearly to the displacement. All response functions can be described in terms of equilibrium expectation values or thermal average of the system's variable in an equilibrium environment.

The extension of the linear response theory to the situation out-of-equilibrium can be realized in terms of the Langevin equation. This is a classical equation of motion,  which incorporates both dissipation and fluctuation effects upon the particle in a random medium. The  dissipation effect is normally accounted for by a friction force as we already described, and in general depends on the past histories of the particle. The competing phenomenon of the fluctuations has the same physical cause. The noise forces mimic the random environment and are correlated over time scales, determined by the typical scales of the medium.  The dissipation term in the Langevin equation represents the steadily transferred system energy into the environment, whereas the information on initial conditions is lost. The noise term, at the same time, works to feed the right amount of fluctuations into the system, so that it evolves toward the equilibrium state.    The fluctuation and dissipation appear to be inseparable and it turns out that they follow the fluctuation-dissipation theorem stating how the magnitudes of the friction and fluctuations are related to each other while in equilibrium.

Such phenomena have a wide range of applications since for most of non-equilibrium systems, their quantitative description relies only on statistical properties of the environmental degrees of freedom in the case that the central limit theorem can be applied.   The underlying assumption to construct the Langevin equation for a Brownian particle is the existence of different timescales in nonequilibrium systems. They are the relaxation time needed for the particle to forget its initial velocity and
thermalize,  and the collision duration time from the kick of the molecule in the medium with a particle, where the random force is correlated.  Normally, the relaxation time scale is larger than that of the collision time scale. The related properties of the random force describing the rapidly fluctuating environment can be  listed by some natural assumptions. a) The force is stochastic over time scale larger than the collision time with a zero mean value; b) the environment is in equilibrium and homogeneous, the stochastic force, which has no spatial dependence, varies chaotically, therefore it is uncorrelated to itself except over the collision time scales; c)  the overall statistical properties of the random force obey the time translational invariance so that their 2-point correlator depends only on the time difference. As a result, the motion consists of series of randomly independent displacements and without any further computation, according to the central limit theorem, it is expected that for the classical Brownian motion they end up to a normal distribution of the displacement variable with growing standard deviation in time as $\sqrt{t}$. Moreover, the average displacement is proportional to the average particle velocity,  which by the equipartition theorem scales with the temperature as $\sqrt{k T}$. Then mean value of the displacement is $\vev{\d x^2}\sim f(\eta) k T t$, where $f(\eta)$ is inversely proportional to the friction coefficient. We notice that the random force correlator does not appear explicitly in the formula, although properties of it has been used to arrive to the above conclusion. The time-integral of the random force correlator can be computed by the Langevin equation, and turns out to depend on the temperature and the diffusion coefficient. This is natural, the hotter the system the more collisions occur.

\textbf{Quantum and Thermal Fluctuations on Quarks and Massive Particles:} It is obvious from the discussion so far, that such phenomena are expected to occur in the strongly coupled systems  in deconfined phase and out of equilibrium.   The most representative case is the interaction of a heavy quark in the quark-gluon plasma. As long as the mass of the quark is much larger than the temperature of the system we expect that its dynamics are described by a diffusion process. From the equipartition theorem the square of the thermal momentum of the quark is proportional to $m_q T\gg T^2$ while the squared momentum transfer of the medium is of order $T^2$, therefore the stochastic analysis is justified.  Equivalently the typical relaxation time scale for a quark, moving in a hot thermal bath, is estimated as $t_{r} \approx 1/(T^2/m_q)$, while the collision time scale is given by $t_c \approx 1/T$. Thus, two time scales are largely separated as $m_q \gg T$ and the quark can be regarded as the system with slow-varying degree of freedom whereas the effects from   fast-varying degrees of freedom of hot thermal bath are effectively summarized as stochastic forces, correlated over the time scale $T$.  In fact by using the approximation of hard thermal loop QCD perturbation theory and heavy mass quark expansion the average energy quark loss has been computed and has been shown that the Langevin equations consistently describe the kinetics of heavy particle in the thermal medium  \cite{Moore:2004tg,vanHees:2004gq,Mustafa:2004dr,Romatschke:2004au}

In quantum environment the fluctuation and dissipation phenomena  occur even at zero temperature due to vacuum fluctuations of the environment fields, originating from the uncertainty principle. Our thermodynamic and statistical arguments above imply that in such cases a diffusion and a type of Brownian motion should exist. In fact it has been shown  by using tools of quantum statistical mechanics that the fluctuation-dissipation theorem is satisfied when  the zero-point energy of the electromagnetic field is taken into account \cite{landausp}. Furthermore, perturbative methods have been developed to investigate such quantum phenomena. The methods consist  of integrating out the environmental degrees of freedom by the use of Feynman-Vernon influence functional  \cite{caldeira1983,Schwinger,Feynman:1963fq}, while modeling it as an infinite number of  simple harmonic oscillators \cite{Grabert:1988yt,Hu:1993qa,Hu:1986jj}. The environment effects then can be classified  according to the effective dependence of their friction on the  frequency as ohmic, supra- or subohmic.   For example, the  effects from the quantized electromagnetic fields  on a point charge in the dipole approximation, corresponding to the supraohmic environment, have been studied  in \cite{Hsiang:2005pz,Hsiang:2007zb,wu3}.

Therefore, taking the analogue of the strongly coupled physics, we expect that the quantum fluctuations of a particle  trajectory  can be induced by its coupling to the gluonic field at zero temperature. The quark interacts with the fluctuations of the gluonic and quarkonic environment which results to its fluctuations. The induced non-uniform motion results to a gluonic radiation back to the medium as an effect of the dissipation process. A valid question to ask is whether such a process always implies a kind of quantum Brownian motion and the fluctuation-dissipation theorem holds, irrelevant of the context of the strongly coupled theory considered. For boost invariant theories, the motion is trivial with a constant speed particle. However for non-boost invariant theories the motion experiences a friction resulting to drain energy into the soft infrared modes of the theory. Depending on the nature of the theory, the particle eventually may travel in an infinite interval,  or simply slow down and stop closed to the region that the motion was initiated.

\textbf{Quark Dynamics in Gauge/Gravity duality:} In this work we provide a universal scheme for quantum and thermal fluctuations implementing our study using holographic methods \cite{adscft1,adscft2}. The test particle chosen as a massive quark represented by an open string hanging from the boundary of the gravitational dual background. Its end-point on the boundary is the position of the particle in the spacetime where the field theory lives. Small perturbations around the string are described by free scalar fields which propagate on the induced worldsheet, therefore the problem reduces to studying the dynamics of two dimensional quantum fields in curved spacetimes. By quantizing the fluctuations of the worldsheet we relate the quantum modes of the string to its boundary endpoint. In the regime of the validity of semiclassical approximation the correlation function of the position of the end-point particle is related to the excitation spectrum of the string world-sheet. One may wonder whether the string fluctuations induce a Brownian motion to its string endpoint representing the particle motion.

The initial studies on the motion of quarks in the quark gluon-plasma in the context of the  gauge/gravity correspondence have been done in \cite{Herzog:2006gh,Gubser:2006bz} where the friction coefficient was determined, while in \cite{CasalderreySolana:2006rq,Gubser:2006nz,CasalderreySolana:2007qw} the integral of the correlator of the random force was obtained and a  review can be found in \cite{CasalderreySolana:2011us}. Later studies have been shown that the   Langevin coefficients of moving particles satisfy a universal inequality $\k_\parallel\ge \k_\perp$ \cite{Giataganas:2013hwa,Gursoy:2010aa}. More particularly, by obtaining the coefficient formulas readily applicable to any holographic background it was shown in \cite{Giataganas:2013hwa} that for isotropic holographic theories the longitudinal Langevin diffusion coefficient along the quark motion is larger compared to that of the transverse direction. This is a universal inequality which however can be reverted/violated in the presence of strongly coupled anisotropies \cite{Giataganas:2013hwa,Giataganas:2013zaa}, in a similar way with the well known shear viscosity over entropy density bound \cite{Kovtun:2004de,Rebhan:2011vd,Jain:2015txa,Giataganas:2017koz}. The fundamental aspects of the Brownian motion for a quark in finite temperature AdS/CFT correspondence were examined in parallel \cite{deBoer:2008gu,Son:2009vu}. In fact the qualitative picture of the nature of the particle becomes clear by following logical steps. The gravity dual theory has a black hole horizon, and it is known that the Hawking radiation induces random motion on the string \cite{Lawrence:1993sg}. Considering the fact that Hawking radiation is consistent with the fluctuation-dissipation theorem \cite{Hawking:1974sw,Gibbons:1976pt,Unruh:1976db,Israel:1976ur} the dissipation of the energy and the corresponding stochastic motion of the particle should be of Brownian nature. This has been  already shown for certain theories and in our work we provide further evidence that this is true for a wide class of holographic theories.

On the other hand for zero temperature fluctuations the literature is less extended. So far it has been found that the particle follows stochastic motion in the case of Lifshitz spacetime \cite{Tong:2012nf} and the hyperscaling violation spacetime \cite{Edalati:2012tc} while earlier works have shown the dragging of the particle even at zero temperature in these theories \cite{Hartnoll:2009ns,Kiritsis:2012ta,Fadafan:2009an}.

In addition to the works already mentioned \cite{Herzog:2006gh}-\cite{Son:2009vu}, the drag/diffusion quark analysis has been used to extract qualitative phenomenological information for the momentum broadening of heavy quarks moving in the quark-gluon plasma also for example in \cite{Rajagopal:2015roa,Fischler:2012ff,Moerman:2016wpv,Dudal:2018rki},  while other works on Lifshitz-related spacetimes include \cite{Yeh:2013mca,Yeh:2015cra,Lee:2017qnr,Roychowdhury:2015mta,Banerjee:2013rca}.

\section{Our Methodology and Findings} \la{sec:impact}

Motivated by the wide range of applicability of the phenomena discussed above we provide a universal study scheme for quantum and thermal fluctuations, the dissipation of the energy and the corresponding Brownian motion of the particle, in a wide class of strongly coupled field theories. The quark fluctuations are studied in zero temperature medium, which are induced by the fluctuations of the zero point energy, and in finite temperature scheme where the thermal bath effects are taken into account. Our gravity dual space has arbitrary scaling dependence on the holographic direction, corresponding to a wide class of dual field theories. The string fluctuations are examined for the generic spacetime metric, while the solution of the string mode expansion turns out to be described by Bessel functions of order $\n$,  which depends on the arbitrary scalings of the metric. The two-point function of the fluctuations exhibits two different late time behaviors, depending purely on the value of the order of Bessel functions, and there is always a crossover region for $\n=1$ which defines the type of growth. Moreover, we find that in one of these regions the correlator is always independent of the mass of the particle.

Having obtained the correlator, we apply a force to the system to take it out of equilibrium and use the linear response theorem justified in the previous paragraphs. By finding the response function  we explicitly show that the fluctuation-dissipation theorem from the bulk perspective is always satisfied irrespective of the choice of the background metric scalings.  The underlying computational reason is that our analytic methods turn out to heavily depend in a compact way on the order of the Bessel functions, incorporating the different theories in a universal way. Knowing the response function we  read the inertial mass and the self energy of the particle. Which of them is dominant in the response, depends exclusively on the order of Bessel function and the critical value is again $\n=1$. A focus of the zero temperature study, is to highlight how central is the role of the Bessel function to the whole analysis, and how compactly the mathematical formulas are expressed in terms of it. The analysis in this section is in agreement with previous works on Lifshitz and hyperscaling violation metrics, which guarantee that our analysis can go through all the way.  

Then we move on to study the Brownian motion on thermal field theories. We work again with the generic class of holographic theories, where in the metric we require the existence of a blackening factor with a single pole. We study the thermal diffusion and compute the boundary fluctuations of the string by bringing their ruling equation to a Schr\"{o}dinger-like form. The approximate string solution in low frequency can be found for the class of the backgrounds by employing a variation of the monodromy patching method \cite{Motl:2003cd,Maldacena:1996ix,Harmark:2007jy}. When having the solution, we employ a force on the system in order to compute the response function and we find it to be always inversely proportional to the frequency and the horizon value of the metric element along the direction that the fluctuation occurs. The diffusion constant turns out to be scaled with the temperature as $T^{2\prt{1-\nu}}$, with the power depending solely on the order of Bessel function, the same for the zero temperature fluctuation. Whether it increases or not depends on the scalings of the metric element incorporated at the order of the  Bessel function.

Then we move to obtain the thermal inertial mass to show that it always receives a contribution due to thermal background compared with the zero temperature one. The correction depends on the temperature, the order $\n$ and the length of the fluctuating string. This is natural since the length of the fluctuating string can be traded with the energy needed to create the string. The next task of our study is to verify the fluctuation-dissipation theorem for the whole class of the thermal backgrounds. Indeed it is satisfied and in the context of the proof, it turns out that the main contributions come from the boundary and horizon string dynamics. This justifies our earliest arguments that the theorem should hold for such fluctuation phenomena due to their statistical nature and the energy conservation. As a final task we show that the random force is a white noise, since its self-correlator is proportional only to the temperature with a power of $T^{2\nu}$.

Having analyzed the string fluctuations, we then consider the study of fluctuation of certain Dp-branes. The problem is formally interesting, while an additional motivation comes from the importance of Brownian motion of cavities. For example perfect or imperfect mirrors in quantized electromagnetic field backgrounds will exhibit a Brownian motion, due to the pressure of the background field and the radiation due to nonuniform acceleration as for example in \cite{Gour:1998my,Wu:2005jr}. Such phenomena find application to the dynamical Casimir effect \cite{Dodonov:2010zza,Milton:2004ya} and  a strongly coupled study would be of great interest. Our Dp-branes considered are of certain rigid type and  live in the d+1-dimensional arbitrary gravity background, the same with the particle study above. We prove that there is one-to-one map between the Dp-brane fluctuation and the string fluctuations. Therefore we are able to determine in a straightforward way for the Dp-branes: the two-point function, the (thermal) response function, the inertial mass, the self energy  and the diffusion constant. All the Dp-brane observables depend on a deformed $\tilde{\n}$ parameter related to the dimension p of the brane and the scalings of the metric elements that the brane extends. The parameter $\tilde{\n}$ is still the order of the Bessel function for the Dp-brane fluctuations. The existence of the mapping of string to brane dynamics is not so surprising. In \cite{Giataganas:2015ksa} it has been shown that the dynamics, of the more involved than in our case  Dp-branes related to $k$-strings, are  expressed in terms of the fundamental string ones. Since the brane configuration there is more involved, an additional background condition was needed for the mapping related to a preserved quantity under the T-dualities, which here is not necessary.

A final comment regarding our approach is on the physical range of the arbitrary power scalings appear in the metric. Our analysis is performed for arbitrary values of them, and in principle may be constrained by the requirement of having physical and stable theories. The null energy condition, constrains the range of parameters ensuring that the gravity theory is non-repulsive. For finite temperature solutions the local thermodynamical stability is ensured by
the requirement of non-positive values of the Hessian matrix of the entropy with the thermodynamical values \cite{Gubser:2000mm}. An additional test of stability is the computation of the entanglement entropy. Unstable regions of the scaling parameter space, will be recognized when the entropy scales parametrically faster than the area \cite{Dong:2012se}.

The organization of the paper is the following. In section \ref{sec:flucpar1} we define the class of holographic theories we use in this paper. Then we provide a generic scheme at the level of actions
for the string fluctuations. We move on to compute the two-point function of the particle fluctuations
and find the crossover regions. In the next section \ref{sec:response1} we compute the response function of the system, from where we read the particle's inertial mass and the self-energy. We also show that the fluctuation-dissipation theorem holds. In section \ref{sec:flucther1}, we consider gravity theory of a finite temperature dual field theory, where we find the string fluctuating solutions with the monodromy matching method. Using this solution we compute the response function from where we read the diffusion constant, the self energy and the thermal corrections to the inertial mass. We also show that the random force is a white noise and find how it scales with the temperature. In the next section \ref{sec:dpfluc1} we show how the Brownian motion of a string can be mapped to a Brownian motion of a rigid Dp-brane and therefore we can map all our string results to the brane ones. This guarantees that the fluctuation-dissipation theorem for the branes is satisfied. Then we move on to demonstrate the application of our generic methods to certain holographic backgrounds in section \ref{sec:appl}. One of these cases are theories with strongly coupled anisotropic dynamics where we find how the Brownian motion is modified compared to the isotropic dynamics. In the final section \ref{sec:conc} we briefly discuss the implications of our results and provide a list of our main equations as an additional aid to the reader. We support our study with three appendices. In the first  we recall the properties of the Bessel functions we have used to bring our results in compact form. In the appendix \ref{app:coor}, we find the coordinate transformation to bring the mode differential equation of the string fluctuations to a Schr\"{o}dinger-like form. In the last appendix we briefly review the general trailing string analysis using the membrane paradigm \cite{Giataganas:2013hwa,Giataganas:2013zaa} and show that in the static limit there is agreement with certain coefficients obtained in this work.

\section{Fluctuation of Massive Particle} \la{sec:flucpar1}

Let us consider massive particles  in the generic  geometry in string frame
\be\la{gen1}
ds^2=-g_{00}(r) dx_0^2+ g_{rr}(r)dr^2 +\sum_{i=1}^{d} g_{ii}(r) dx_i^2 ~,
\ee
with $\lim_{r\to \infty} g_{ii}(r)=\infty$, such that the boundary is at $r=\infty$. $d$ are the space dimensions and the metric is diagonal.

The massive particles are represented in holography by strings, initiating from a point in the holographic direction $r=r_b$ located close to the boundary, at a probe D-brane that fills the spacetime. The string then extends to the IR $r\rightarrow r_h$, where $r_h\neq 0$ is the horizon of the black hole when the considered spacetime \eq{gen1} is dual to a quantum field theory of finite temperature, or $r_h=0$ when the field theory under study is of zero temperature and the string terminates at the deep IR.

Let us work in the radial gauge and without loss of generality in the linear region, consider the equation of motion
for the $x_1$ direction. In the following, all the results are generalized in a straightforward way
by replacing $x_1$ (and the corresponding metric element $g_{11}$) to the rest of the directions.
Notice that when there are potential anisotropies in the space-time, the quantum fluctuations will depend on the directions considered. The way we parametrize the string is
\be
r=\s~,\qquad  x_1=x_1(\t,\s)
\ee
and the coordinate time matches with the string world-sheet spacetime.
The string ending on a D-brane has the Nambu-Goto (NG) action
\be\la{actiona1}
S=-\frac{1}{2 \pi \a'} \int d\s d\t\sqrt{-\prt{g_{00}+g_{11} \dx_1^2} \prt{g_{rr}+g_{11} x_1'^2}}~,
\ee
with
\be \la{Nbc}
x_1'(r_b)=0~,
\ee
the Neumann boundary conditions imposed at the brane at $r=r_b$. A solution to the equations of motion is $x_1=c$, representing a straight string hanging from the brane to the bulk. The energy to create this string is simply
\be\la{energy}
E=\frac{1}{2\pi \a'}\int_{r_h}^{r_b} dr \sqrt{-g_{00} g_{rr}}~,
\ee
where $r_h=0$ when we are interested on the study of a particle in a zero temperature heat bath. In a non-relativistic theory the energy of the string is not the same as its inertial mass, and we provide a generic formula for their relation in a following section.

Since we know the generic solution of the string we can study the (quantum) fluctuation of its end point, by computing the two point function $\vev{X_1(t) X_1(0)}$ where $X_1(t):=\d x_1(t,r_b)$.
Without loss of generality let us consider the small fluctuations of the endpoint of the string at the brane around $x_1(t,r_b)=0$. The Nambu-Goto action of the fluctuations $\d x(t,r)$ turns out to be
\be\la{actionsec}
S=c-\frac{1}{4 \pi \a'} \int d\s d\t  \prt{-\frac{g_{11} \sqrt{-g_{00}}}{\sqrt{g_{rr}}}\d x_1'^2+ \frac{g_{11}\sqrt{g_{rr}}}{\sqrt{-g_{00}}}\d \dx_1^2}~.
\ee
The equation of motion reads
\be\la{eqfluc1}
\frac{\pp}{\pp r}\prt{\frac{g_{11} \sqrt{-g_{00}}}{\sqrt{g_{rr}}}\d x_1'}-\frac{g_{11}\sqrt{g_{rr}}}{\sqrt{g_{00}}}\d \ddot{x}_1^2=0~.
\ee
We can Fourier decompose the most general solution   to the above linear equation as an integral over the frequencies
\be\la{flucgen}
\d x_1(t,r)=\int_0^\infty d\om h_\omega (r) \prt{\a(\omega)e^{-i\omega \t}+\a(\omega)^\dagger e^{i\omega \t}}~,
\ee
where $\a(\omega)^\dagger, \a(\omega)$ being the creation and annihilation operators.
The mode equation takes the form
\be\la{eqmodes1}
\frac{\pp}{\pp r}\prt{\frac{g_{11} \sqrt{-g_{00}}}{g_{rr}}h_\om(r)'}+\om^2\frac{g_{11}\sqrt{g_{rr}}}{\sqrt{g_{00}}}h_{\om}(r)=0~.
\ee
In order to proceed finding a solution we need to specify the generic form of the metric elements.

\subsection{Quantum Fluctuation and Two-point Functions in General  Theories}\label{section:QFTT}

To proceed and solve the generic differential equation \eq{eqmodes1}, we need to consider a generic class of theories with dual backgrounds as in \eq{gen1}. Let us take metrics with polynomial metric elements as
\be\la{polymetric}
g_{00}=r^{a_0} f(r)~,\qquad g_{rr}=\frac{1}{r^{a_u} f(r)}~,\qquad g_{ii}=r^{a_i}~,
\ee
where the $i$ indices count the spatial directions and $a_i$ are constant powers. The class of the dual field theories to the above metric is  wide, to mention some of them, all the non-relativistic theories, including the hyperscaling Lifshitz violating ones \cite{Kachru:2008yh,Dong:2012se,Narayan:2012hk}
or the anisotropic theories \cite{Azeyanagi:2009pr,Mateos:2011ix,Mateos:2011tv,Giataganas:2017koz,Jain:2014vka,Donos:2016zpf} and several others. Our string fluctuation analysis in certain cases and limits is even applicable for  backgrounds with UV and IR asymptotics of \eq{polymetric} like particular RG flows. In this sense, we establish a complete treatment on the quantum fluctuations in (non-)relativistic systems, working in a large class of physical theories.

In this section let us consider a particle in a zero temperature environment, setting
\be
f(r)=1,~\quad r_h=0~.
\ee
A comment regarding the metric \eq{polymetric} at zero temperature is in order. One may rescale $r$ to fix one of the constant scalings, say $a_u$. We choose not to fix it for three reasons: i) we will see that our methods and results are formulated in terms of a constant $\n$ depending on all the scalings of the background and not independently on each scale, so the presentation would remain the same even if we fix one scaling; ii) for reasons of convenience, to provide generic results applicable directly to any gravity background without the need of any coordinate transformation; iii) to make a direct connection to the generalization of the finite temperature method.  The fact that  the hyperscaling violation analysis (with two independent scaling parameters as here) has been done at zero temperature \cite{Edalati:2012tc}  \footnote{In fact even the Lifshitz analysis with one scaling parameter \cite{Tong:2012nf} could guarantee that, since we show below the solutions of the fluctuations for all the backgrounds \eq{polymetric} belong to the same unique class, and the analysis depends on certain common characteristics of the class of solutions.} guarantees that our generic methods at zero temperature will go through all the way. Our theories are more generic than the hyperscaling ones, including for example the anisotropic theories, and theories with  different stability and physical ranges compared to hyperscaling theories, allowing additional features to be seen. Moreover, in our notation where each scaling is labeled uniquely, we can track explicitly how the different metric elements affect the observables, and this is crucial to deduce more generic formulas. Based on that, a key observation of this section is, how central to the analysis the order of the Bessel function is, where most of the stochastic observables depend exclusively on it.  This is true even at finite temperature, where the zero temperature study of this section is essential to make the connection. Finally, our methodology is crucial for the mapping of the string to brane fluctuations we study in coming sections.  We comment further on these issues later.

The second order differential equation for the modes \eq{eqmodes1} become
\be\la{eqmodes2}
\frac{\pp}{\pp r}\prt{r^{a_1+\frac{a_0+a_u}{2}} h_\om(r)'}+\om^2r^{a_1-\frac{a_0+a_u}{2}}h_{\om}(r)=0~,
\ee
and has solutions of Bessel type, which after some manipulation can be written as
\be\la{solmodes1}
h_\om(r)=r^{-\nu \k} A_\om \prtt{J_\nu\prt{\om \trr}+B_\om Y_\nu\prt{\om \trr}}~,
\ee
where
\be\la{def1}
\nu:=\frac{a_0+2 a_1+a_u-2}{2\prt{a_0+a_u-2}}~,\qquad \trr:=\frac{2 r^{\frac{1}{2}\prt{2-\a_0-\a_u}}}{a_0+a_u-2}=\frac{r^{-\k}}{\k}~,\qquad \k:= \frac{a_0+a_u}{2}-1~
\ee
and $J_\nu(\trr),Y_\nu(\trr)$ are the Bessel functions of first and second kind. Notice from the above definitions that $\k\prt{2\n-1}=a_1$. To obtain the integration constants we need to look at the canonical commutation relations of the theory \eq{actionsec}. The conditions for the operators \eq{flucgen} read
\be\la{commut}
[\a_i,\a_j] =[\a^\dagger_i,\a^\dagger_j ]=0~, \qquad [\a_i,\a^\dagger_j]=\d_{ij}
\ee
and are translated to a basis of normalized functions $u (t,r)=h_\om(r) e^{-i\om t}$ satisfying the equation \eq{eqfluc1} as
\be\la{uconditions}
\prt{u_i,u_j}=-\prt{u^\star_i,u^\star_j}=\d_{ij}~,\qquad \prt{u_i,u^\star_j}=0~.
\ee
Here the inner product is of Klein-Gordon type \cite{Birrelldavies}
\be\la{inner1}
\prt{u_i,u_j}_\Sigma=-\frac{i}{2 \pi \a'}\int_\S \sqrt{h}~ n^\m g_{ij}\prt{u^i\pp_\m u^j{}^\star-\pp_\m u^i u^j{}^\star}~,
\ee
and it is independent of the choice of the hypersurface $\S$. Therefore, we choose a constant time slice surface  with a unit normal vector field being $n^t=1/\sqrt{-g_{00}}$.

From the canonical commutation relations we specify the constants of motion.  The inner product is written as
\bea\nn
(u_1,u_2)=\frac{\k e^{-i\prt{\om-\tom}t}}{2 \pi\a'} \prt{\om-\tom}A_\om A_\tom
\int d\trr~ \trr \prt{J_\n\prt{ \om\trr}+B_\om Y_\nu \prt{\om\trr}}\cdot
\prt{J_\n\prt{\om\trr}+B_\om Y_\nu \prt{\om\trr}}~,
\eea
where $\n$ and $\trr$ are defined in \eq{def1} and depend on the scalings of the metric.
Using the properties of the Bessel functions presented in Appendix \ref{appendix:bes} and the ones of the Dirac $\delta$ functions, after some algebra we obtain a surprisingly compact result for the first constant
\be
 A_\om=\sqrt{\frac{  \pi \a'}{|\k|\prt{1+B_\om^2}}}~,
\ee
where the absolute value is generated by the delta function properties. $A_\om$ depends on the scaling behavior of time and radial metric elements in a simple way.

The constant $B_\om$ is determined by the use of the Neumann boundary conditions \eq{Nbc}. To do that we use the properties of the derivatives of Bessel functions of Appendix \ref{appendix:bes}, which are taken with the respect to $\om \trr$ using a coordinate redefinition
\be
\frac{\pp (\om \trr)}{\pp r} \frac{\pp J_\n(\om \trr)}{\pp (\om \trr)}=-\omega r^{-\k-1}\frac{\pp J_\n(\om \trr)}{\pp (\om \trr)}~.
\ee
The second constant turns out to be equal to
\be\la{solb}
B_\omega=- \frac{J_{\n-1}\prt{\omega \trr_b}}{Y_{\n-1}\prt{ \omega\trr_b}}~,
\ee
where $\n$ and $\trr$ have been defined in \eq{def1} and therefore $B_\om$ depends on   the scalings of the metric  background.

\begin{figure}
\begin{minipage}[ht]{0.5\textwidth}
\begin{flushleft}
\centerline{\includegraphics[width=90mm]{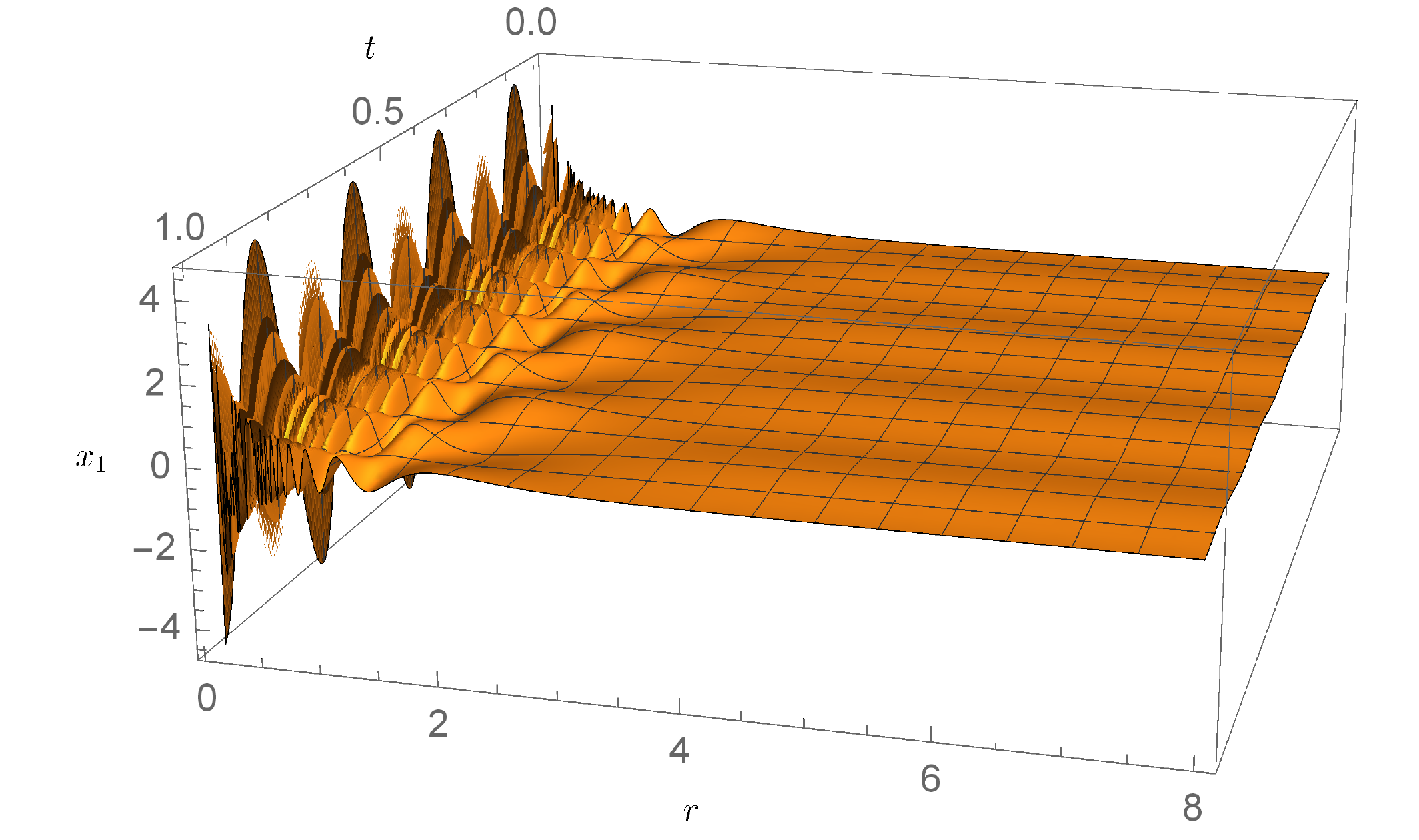}}
\caption{\small{The real part of the time dependent fluctuations $x_1(t,r)=e^{-i \om t} h_\om(r)$ for a fixed frequency and scalings of the background (taken as Lifshitz for the plot). Notice the large magnitude and density of excitations in the deep IR-small $r$, which result to the corresponding Brownian motion on the boundary-large $r$.}}
\label{fig:xt1}
\end{flushleft}
\end{minipage}
\hspace{0.3cm}
\begin{minipage}[ht]{0.5\textwidth}
\begin{flushleft}
\centerline{\includegraphics[width=65mm ]{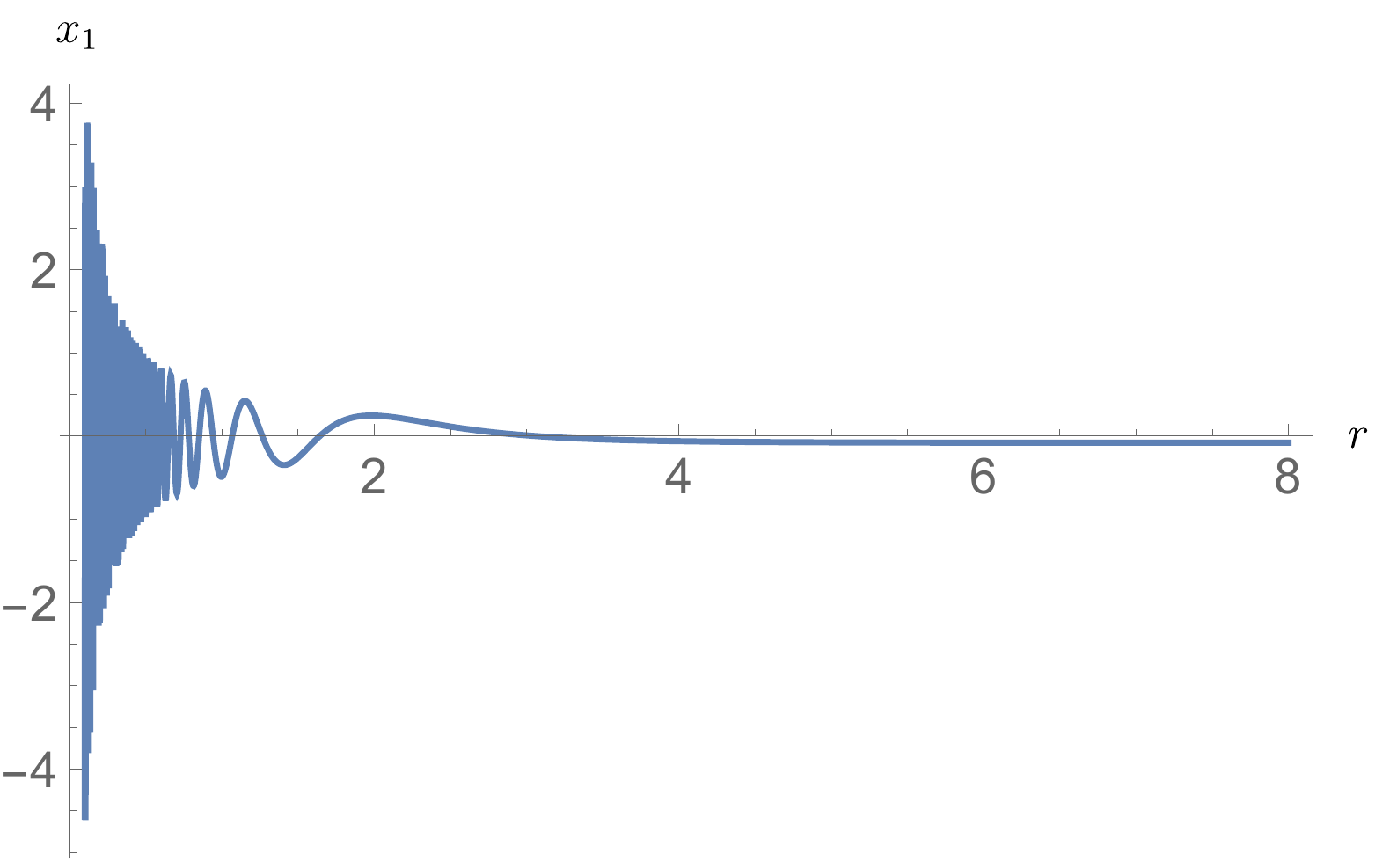}}
\caption{\small{The real part of the  fluctuations  a fixed frequency, time and scalings of the background. The large magnitude and frequency of excitations in the deep IR, weaken as we approach the boundary.} }
\label{fig:xt2}\vspace{.0cm}
\end{flushleft}
\end{minipage}
\end{figure}

Collecting all our results the solution \eq{solmodes1} becomes
\be\la{solmodes2}
h_\om(r)=r^{-\n\k} \sqrt{\frac{ \pi \a'}{|\k|\prt{1+B_\om^2}}} \prtt{J_\nu\prt{\om \trr}+B_\om Y_\nu\prt{\om \trr}}~,
\ee
where the constant $B_\omega$ given by \eq{solb} and the solutions have a naturally expected behavior, shown in Figures \ref{fig:xt1} and  \ref{fig:xt2}. Thanks to the fact that the differential equation of the string modes in the generic spacetime has solutions of Bessel type, and by taking full advantage of the general properties of the Bessel functions we end up with such a compact result. The modes turn out  to depend on the scaling powers of the metric elements where the string fluctuates in a simple way. The analysis  is quite powerful, since for any background of the type \eq{polymetric},  the solution of the modes is given by \eq{solmodes2}.

We now are ready to compute the two-point function of the end-point of the string. To  compute $\vev{X_1(t) X_1(0)}$ we make use of the commutation relations \eq{commut} and we define as usual the vacuum state such that $a(\om)\ket{0}=0$. So by applying the mode solution \eq{flucgen}, \eq{solmodes2} we get
\be
\vev{X_1(t) X_1(0)}=\int_0^\infty d\om \tom \d\prt{\om-\tom} e^{-i\om t} h_\om(r) h_\tom(r)~.
\ee
Making use of the Bessel function properties of different kind and consecutive neighbors of Appendix \ref{appendix:bes}, we can show that
\be
h_\om(r) h_\om(r)= \frac{4 \a'}{\pi}\frac{r^{-a_1}}{\om^2  |\trr|} \prtt{J_{\nu-1}\prt{\om \trr}^2+Y_{\nu-1}\prt{\om \trr}^2}^{-1}~,
\ee
giving for the two-point function
\bea \nn
\vev{X_1(t) X_1(0)}&=&\int_0^\infty \frac{d\om}{2\pi} e^{-i\om t}\vev{X_1(\om) X_1(0)}
\\\hspace{1.4cm}&&= \int_0^\infty d\om e^{-i\om t} \frac{8 \a' r^{-a_1}  }{\om^2 |\trr_b|} \prtt{J_{\nu-1}\prt{\om \trr_b}^2+Y_{\nu-1}\prt{\om \trr_b}^2}^{-1}~,\la{twopdef}
\eea
where we remind that the $\nu$ and $\trr$ are defined in the \eq{def1}. It is remarkable that all the information of the background is incorporated in the two point function in a simple way.
To integrate the function let us look at the low frequency regime defined as
\be\la{smallo}
\om\ll \trr_b =\frac{r_b^{-\k}}{\k}~.
\ee
In this region  by  expanding the Bessel functions we obtain
\be
\vev{X_1(\om) X_1(0)}\sim \frac{\trr_b r_b^{-a_1}}{\prt{\trr_b \om}^2 \prt{c_0+c_1 (\trr_b \om)^{2-2\n }+c_2 (\trr_b \om)^{ 2\n -2}+c_3 (\trr_b \om)^{ 2\n }}}~,
\ee
to find that the leading order depends on the asymptotics of the geometry as
\be
\vev{X_1(\om) X_1(0)}\sim
\begin{cases}
r_b^{4\k(1-\n)} ~\om^{  2\n-4 } ~,\quad \rm{when}\quad \n \ge 1~,\\
r_b^0 ~\om^{ -2\n}~,\quad \rm{when}\quad  \n \le 1~,
\end{cases}
\ee
where $\n , \trr$ are defined in \eq{def1}. Before we elaborate further on our results let us express the
the previous equation in terms of the energy of the string. Applying our expression \eq{energy} we get
\be
E=\frac{1}{2 \pi\a'}  \prt{a_0-\k} r_b^{a_0-\k}~.
\ee
Therefore the integrand of the two-point function in the low frequency limit in terms of the energy of the string can then be expressed as
\be\la{twopointen}
\vev{X_1(\om) X_1(0)}\sim
\begin{cases}
E^\frac{4\k\prt{1-\n}}{ \prt{a_0-\k}} ~\om^{ 2\n-4}~,\quad \rm{when}\quad \n \ge 1 ~,\\
E^0 ~\om^{ -2\n}~,\quad \rm{when}\quad  \n \le 1~,
\end{cases}
\ee
The inequalities for $\nu$ of \eq{twopointen} are translated to  inequalities of the scalings $a_i$ of the metric elements. By making the sensible assumption that at late times the dominant contribution to the two-point function comes from the low frequency limit of \eq{twopointen}, we obtain
 \be\la{twopointf}
\vev{X_1(t) X_1(0)}\sim
\begin{cases}
E^\frac{4\k\prt{1-\n}}{ \prt{a_0-\k}} ~|t|^{3-2\n}~,\quad \rm{when}\quad \n \ge 1~, \\
E^0 ~|t|^{ 2\n-1}~,\quad \rm{when}\quad  \n \le 1~,
\end{cases}
\ee
where we remind the reader that $\n $ is defined in \eq{def1} and depends on the scaling factors of the metric and is the order of the Bessel functions in the fluctuations of the string. After an involved computation the two point function behavior depends mainly on a single parameter, the order of the Bessel function determined  by the dual metric.

We observe that in any gravity dual theory belonging in the class of metrics considered, the two-point function at late times has remarkably different behaviors depending on the values of $\n$, while it has also some universal characteristics something that is guaranteed already by \cite{Tong:2012nf,Edalati:2012tc}. There is a region in the parameter space of the class of the geometries, defined by $ \n \le 1$, where the long-time correlation of the particle is independent of the mass. We further see that there is always a crossover in the behavior for
\be\la{crossover}
\n=1\Leftrightarrow 2a_1=a_u+a_0-2~,
\ee
where the two-point function grows linearly with $t$. In the positive $\nu$ region this is the maximum rate of growth of the two point function.

Notice that in both branches of \eq{twopointf} there are subregions of the parametric space $\n$ that excluded by the requirement for non-repulsive gravity, i.e satisfaction of the null energy conditions of \eq{twopointf}. It can be proven that it is always possible to satisfy the null energy condition (NEC)  for subregions of $0<\n \le 1$ and $\n \ge 1$. Moreover, we can show that there exist a negative $\n$ subregion that is also satisfying the NEC, where we need to be more careful since it is known that theories in this subclass may have instabilities \footnote{For example, when applying our analysis to hyperscaling violation theories, the region $\th>d$ for the dynamical exponent belongs to the $\n<0$ region, and although consistent with the NEC,   there are known instabilities on the gravity side \cite{Dong:2012se}.}. However this is not certain to be the case for all non-relativistic theories, e.g. the anisotropic ones may still be stable in such regions.

The minimum rate of growth can be realized when $\n=3/2$ and $\n=1/2$ which result to a logarithmic behavior of the two point function. The AdS gravity dual theory belongs to this sub-class of backgrounds.

Our work shows of the existence of the two regions for field theories realized in holography described by the class of the backgrounds considered here, as has been already shown in \cite{Tong:2012nf,Edalati:2012tc}. In particular the analysis on the zero temperature hyperscaling violation metric which has two free parameters \cite{Edalati:2012tc}, implies few of the findings on this section. Our analysis offers additional information and accommodates more theories and several new features, as we saw above, where advantages of our approach have been described already at the beginning of this subsection. Moreover, it is intriguing that our analysis boils down on simple dependency on the order of the Bessel functions ruling the fluctuations. Such universal treatments depending on the order of Bessel function have been observed before in holography while studying completely different topics of non-integrability and chaos \cite{Giataganas:2014hma}. 

\section{Dissipation and the Response Function}   \la{sec:response1}

Having analyzed the quantum fluctuations let us study the dissipation of the particle's momentum. Let us consider an external force $F(t)= E e^{-i\om t} F(\om)$ acting on the particle, i.e. to the end-point of the string by the existence of an electric field $E$ on the D-brane. The linear response of the system due to the force reads
\be\la{response}
\vev{X_1(\om)}=\chi(\om) F(\om)~,
\ee
where $\chi(\om)$ is the admittance quantifying the response to the system when the external force is added and is practically the retarded Green's function. The force on the string results to an additional boundary term in the string action as
\be
S=S_{NG}+S_{EM}~,\qquad S_{EM}=\int dt \prt{A_t+\vec{A}\cdot \vec{x}}~,
\ee
where without loss of generality and as in the previous section, we can  work with the direction $x_1$, considering the fluctuation of the field along this direction. All our results are generalized for motion along the other directions in a straightforward way, by replacing $x_1$ (and the corresponding metric element $g_{11}$) in the following expressions.

The Neumann boundary condition involves the external force and is computed by the derivative $\pp  S_{NG}/\pp \d x'$ of to give
\be \la{boundf}
\frac{1}{2\pi\a'} \frac{g_{11} \sqrt{-g_{00}}}{\sqrt{g_{uu}}}\d x_1'(r_b)=F(t)~.
\ee
Since the bulk dynamics are not modified, the solution to the equation of motion will be similar to the previous section, but with different boundary conditions and therefore different way that the Bessel functions are combined. In the IR we consider ingoing boundary conditions, appropriate to the computation of the retarded Green function.

To find  the solutions we consider the metric \eq{polymetric} with  generic scaling metric elements \eq{gen1}. The IR dynamics of the fluctuations, can be examined using the   tortoise coordinate transformation
\be
r_\star:=\trr=\frac{r^{-\k}}{\k}~,
\ee
where $\trr$ and $\k$  have been already defined in \eq{def1} and appear in the Bessel function argument of the string fluctuation in the previous section. The string action is
\be
S\sim -\frac{1}{4\pi\a'}\int dt d\trr \prt{\k}^{-\frac{a_1}{\k}} \trr^{-\frac{a_1}{\k}}\prt{ \d x_1'^2-\d \dx_1^2}~,
\ee
giving the equation of motion
\be\la{wave1}
\d x_1''-\frac{a_1}{\k \trr} \d x_1'-\d\ddot{x}_1=0~.
\ee
Taking the IR limit $r\rightarrow 0$ we have $\trr \rightarrow \infty$ by assuming $\k>0$, we find that the \eq{wave1} behaves like the equation in the flat space with solutions given by
\bea
 x^{(in)}(t,r)\sim e^{-i \om \prt{t+\trr}}~,\quad\mbox{where}\quad x^{(out)}(t,r)\sim e^{-i \om \prt{t-\trr}}~.
\eea
To satisfy the ingoing boundary condition we select the first Hankel function: $H:=H^{(1)}=J+iY$, to obtain the following solution written in a very compact form
\be\la{solmodes3}
\d x_1(t,r)=e^{-i\om t} g_\om(r)~,\qquad  g_\om(r)=r^{-\n\k}   H_\n\prt{\om \trr}~,
\ee
using the definitions \eq{def1}. The boundary condition \eq{boundf} can be computed using the properties of the Hankel functions of Appendix \ref{appendix:bes}, and their asymptotics assuming $\k>0$. We find that
\be
F(t)=\frac{e^{-i\om t}}{2\pi\a'} A_\om \om r_b^{\k\prt{\n-1}} H_{\n-1}(\om \trr_b)~.
\ee
The response function then can be obtained from \eq{response} to give
\be\la{response1}
\chi(\omega) =\frac{2\pi\a'}{\om} r_b^{-a_1} \frac{H_{\n }(\om \trr_b)}{H_{\n-1}(\om \trr_b)}~.
\ee
Notice the elegant and compact way that the above expression takes, incorporating all the elements of the system in the definition of $\n$ and $\trr$.

Let us confirm that there are no constrains on the validity of the fluctuation-dissipation theorem. The theorem states that the correlation function is related to the imaginary part of the response function and certain distributions to take into account the dissipation due to thermal noise and quantum noise, as
\be\la{theoremfd}
\vev{X_1(\omega)X_1(0)}=2 \prt{n_B(\om)+1} \mbox{Im}\chi(\om)~.
\ee
$n_B(\om)=\prt{e^{\b\om}-1}^{-1}$~ is the Bose-Einstein distribution to capture the thermal noise effect and the unit corresponds to the zero temperature quantum noise contribution.

To verify the theorem we need to bring the Bessel function of the imaginary part of the responde function to the form appearing in the correlation function \eq{twopdef}. Using the identities of Appendix \ref{appendix:bes} mixing the first and second kind Bessel function and relating the nearest neighbors we get
\be
 \mbox{Im}\chi(\om)=\frac{4 \a'}{ \omega^2}\frac{r_b^{-a_1}}{\trr_b}~\prt{J_{\nu-1}^2\prt{\om \trr_b}+Y_{\nu-1}^2\prt{\om \trr_b}}^{-1}~,
\ee
where $r_b^{-a_1}/\trr_b=\k r_b^{\k-a_1}$ using the definition of \eq{def1}.
Comparing it with \eq{twopdef} we find that is proportional to the correlation function:
\be\la{flucdis}
\mbox{Im}\chi(\om)= \frac{1}{2}\vev{X_1(\omega)X_1(0)}~,
\ee
confirming the zero temperature fluctuation-dissipation theorem for string propagating in generic space-times at zero temperature. The confirmation itself at zero temperature is implied from the hyperscaling  violation analysis of \cite{Edalati:2012tc}. The methodology developed here with the generic notation is also useful to extend the study later at the finite temperature theory. There are several works verifying the fluctuation-dissipation theorem in different contexts of AdS black hole and Lifshitz hyperscaling violation theories  \cite{deBoer:2008gu,Son:2009vu,CaronHuot:2011dr,Sonner:2012if,Tong:2012nf}.

Let us look at the low-frequency expansion defined by \eq{smallo}, of the response function focusing mostly on positive $\n$ values. It takes asymptotically the form for non-integer $\n$
\be
\chi(\om)\sim -c_1\prtt{ \frac{r^{2\k\prt{\n-1}}} {2 \k \prt{\n-1}}\prt{i\om}^2 +\cO(\o^4) +\frac{1-i\tan\prtt{\prt{\n-\frac{1}{2}}\pi}}{\prt{i\prt{2 \k}^{2\n-1}\G\prt{\nu}^2}} \pi \prt{-i\omega}^{2\n}+\cO(\o^{2\n+2})}^{-1}~.
\ee
The coefficient proportional to the $\prt{i\om}^2$ is interpreted as the inertial mass $m$ and the coefficient proportional to the $\prt{-i\omega}^{2\n}$ is the self-energy $\g$ of the particle
\be\la{massin}
m=\frac{r_b^{2\k\prt{\n-1}}} {2 \k \prt{\n-1}}~,\qquad \g=\frac{1-i\tan\prtt{\prt{\n-\frac{1}{2}}\pi}}{\prt{ \prt{2 i \k}^{2\n-1}\G\prt{\nu}^2}} \pi ~.
\ee
The competition of dominance between the terms of the denominator is specified
by the value of the order of the Bessel function $\n$ related to the metric scalings as \eq{def1}. For $\n<1$ the self-energy dominates over the inertial mass at low frequencies. It comes with no surprise that for similar values of $\n$, we observe the change of the behavior of the two-point function \eq{twopointf}, while for the same region $\n<1$ we found its independence of the mass. We also notice that there is an allowed region for the scaling parameters where the mass becomes negative or diverges.
In the next section we see how the thermal mass is corrected in the same generic background and how the self energy is modified. 

\section{Thermal Diffusion} \la{sec:flucther1}

Let us now turn on a heat bath and consider the particle fluctuations in a finite temperature quantum field theory. We will study the thermal fluctuations of the endpoint of the string in the context of the Brownian motion. The gravity dual theory given by \eq{gen1} and \eq{polymetric}, contains a black hole and therefore in this section $r_h\neq0$ with
\be\la{black1}
f(r):=1-\frac{r_h^{a_f}}{r^{a_f}} ~,
\ee
where $a_f$ is a constant. The form of the blackening factor is such that is has a single zero at the position of the horizon $r_h$. The temperature of the heat bath of the quantum field theory is given by
\be\la{temperature}
T=\frac{a_f}{4\pi}r_h^{\k}~,
\ee
where $\k$ is defined in \eq{def1}. Most of the following analysis holds for more general forms of blackening factor \eq{black1}. It depends on certain asymptotic  and near horizon behavior and is applicable even of RG gravity flow solutions at certain limits. We elaborate further on that below. The string dynamics follow the generic equations \eq{actionsec} and \eq{eqfluc1}, derived in the previous sections, by applying it to the black hole background. Thinking in the same way we can expand the solution in Fourier modes as $\d x_1(t,r)=e^{-i\om t} h_{\om}(r)$ to obtain
\be\la{eqmodes22}
\frac{\pp}{\pp r}\prt{r^{a_1+\frac{a_0+a_u}{2}} f(r) h_\om(r)'}+\om^2\frac{r^{a_1-\frac{a_0+a_u}{2}}}{f(r)} h_{\om}(r)=0~.
\ee
Let us use the coordinate transformation presented in Appendix \ref{app:coor} to bring the mode equation
\eq{eqmodes22} to the Schr\"{o}dinger-like form
\be\la{schr}
\frac{\pp^2 y}{\pp r_\star^2}+\prt{\om^2-V(r)}y=0~,
\ee
where
\bea\la{rstar}
&&y=h_\om(r) r^\frac{a_1}{2}~,\qquad r_\star= -\frac{r_h^{-\k}}{a_f} ~B_{\prt{\frac{r_h}{r}}^{a_f}}\prtt{\frac{\k}{ a_f},0} ~,\\\la{yeq}
&&V(r)=\frac{a_1}{2} r^{2\k} f(r)\prt{\prt{a_0+a_u+\ff{a_1}{2}-1}f(r)+r f'(r) }~,
\eea
where $B_z\prtt{a,b}$ is the incomplete beta function. The \eq{schr} does not have analytic solution for the whole range of the string for arbitrary background scaling parameters. However, it is possible to find an analytic solutions at different regions
of space for arbitrary scalings by applying the monodromy patching procedure \cite{Motl:2003cd,Maldacena:1996ix,Harmark:2007jy,deBoer:2008gu,Tong:2012nf}.

We look for an analytic solution near the boundary which is ingoing at the horizon of the black hole. To do that we need to consider three different regions that have overlap to each other and mediate the properties of the near horizon solution  to its properties near the boundary. The regions and the method we follow is described below:\newline
I) The near horizon region with
\be \la{r1}
r\sim r_h\qquad\mbox{and}\qquad V(r)\ll \om^2 \Leftrightarrow f(r)\ll \om~.
\ee
Here the potential of the \eq{schr} can be neglected while obtaining the solution approximately.\newline
II) The intermediate region between the horizon and the boundary where
\be \la{r2}
V(r)\gg \om^2 \Leftrightarrow f(r)\gg \om~.
\ee
This solution serves as the transmitter propagator between the boundary and the bulk. Therefore, to make the connection we need to consider the two limits of the same solutions approaching the regions, but still respecting the initial limit \eq{r2}. We will find two solutions for $r\sim r_h$ and $r \rightarrow \infty$. \newline
III) The near to the boundary solution $r\rightarrow \infty \Leftrightarrow r\gg r_h$.

Moreover, we work in the low frequency limit. Starting from region I and neglecting the term of the potential we obtain
\be
y_A(r)=c_1 e^{-i\om r_\star}~,
\ee
where we have kept the purely ingoing solution. The solution $r_\star$ \eq{rstar} close to the horizon becomes
\be
r_\star=\frac{r_h^{-\k}}{a_f} \log\prt{\frac{r}{r_h}-1} ~
\ee
and using \eq{yeq} we finally get
\be\la{aa1}
h_{Ah}(r)=c_1\prt{1- \frac{i\om r_h^{-\k}}{a_f} \log\prt{\frac{r}{r_h}-1}}~.
\ee
Getting to the region II, we neglect the $\om^2$ constant term as the subleading one. We may use now \eq{eqmodes22} where we have the total derivative
\be
h'(r)=\frac{c_4}{f(r) r^{1+2\k\n}}~,
\ee
with solution
\be
h_B(r)=c_3+ c_4 \frac{  r_h^{-2\k\n}}{a_f}   B_{\prt{\frac{r_h}{r}}^{a_f}} \prtt{\frac{2\k\n}{ a_f},0}~.
\ee
To apply the patching method successfully we need to find the near horizon solution
\be \la{bh1}
h_{Bh}(r)=c_3+c_4 c_0+\frac{c_4}{a_f} r_h^{-2\k\n} \log\prt{\frac{r}{r_h}-1}~,
\ee
where $c_0$ is an integration constant depending  on scalings of the metric through the digamma function. Close to the boundary the solution reads
\be\la{bb1}
h_{Bb}(r)=c_3-\frac{  c_4 }{2\k\n} r^{-2\k\n}~.
\ee
Finally to make the connection with the boundary we study the region III, and solve \eq{eqmodes22} close to the boundary to get
\be\la{c1}
h_{Cb}(r)=c_5+c_6 r^{-2\k\n}\prt{\frac{\om ~ \mbox{sign}\prt{1+2\k}}{2\k}}^{2\n}~.
\ee
Let us start the patching of the solutions $h_{Ah}(r)$ \eq{aa1} and $h_{Bh}(r)$  \eq{bh1} near the horizon to obtain
\be\la{eq31}
c_3=c_1-c_0 c_4~,\qquad c_4=-i\om  c_1 r_h^{a_1}~.
\ee
Doing the same near the boundary for the solutions $h_{Bb}$ \eq{bb1} and $h_{Cb}$ \eq{c1} we get
\be\la{eq53}
c_5=c_3~,\qquad c_6 =-\frac{ c_4}{2\k \n}\prt{ \frac{2\k}{\om ~ \mbox{sign}\prt{1+2\k}}}^{2\n}~.
\ee
Combining the results from boundary to the bulk \eq{eq31} and \eq{eq53} we get
\be\la{comb3153}
c_5=c_1\prt{1+ i c_0 \om  r_h^{a_1} }~,\qquad c_6 = \frac{  i\om  r_h^{a_1}c_1}{2\k\n} \prt{ \frac{2\k}{\om ~ \mbox{sign}\prt{1+2\k}}}^{2\n}~,
\ee
to obtain the solution near the boundary
\be\la{hcresult}
h_\om(r)=c_1\prt{1+i\omega c_0 r_h^{a_1}+\frac{i \om r_h^{a_1}  }{2\k\n}r^{-2\k\n}}~.
\ee
This is the behavior of the string fluctuations at the boundary, and we expect it be of diffusive nature.

\subsection{Diffusion Constant}

The expectation value in the canonical ensemble is related to the diffusion constant $D$  as $\vev{:\prt{X_1(t)-X_1(0)}^2:}=2\prt{d-1} D t$, where the normal ordering $:$ serves to keep the finite fluctuations due to the thermal bath. The diffusion constant is related to the response function and the temperature of the heat bath as
\be \la{diffusion1}
D=T \lim_{\om\rightarrow 0}\prt{-i~\om \chi(\om)}~,
\ee
where the linear response is given in terms of the external force \eq{response}, which in its own turn is given in terms of the derivatives of the boundary string fluctuations by \eq{boundf}. The response function   using \eq{hcresult} turns out to be
\be\la{chifint}
\chi(\om)=\frac{2\pi \a'}{-i\om r_h^{a_1}}~.
\ee
It is inversely proportional to the spatial metric element at the horizon along the direction of the fluctuations. In the low frequency limit we propose that this should be true for any arbitrary background metric reading
\be\la{responsemet}
\chi(\om)=\frac{2\pi \a'}{-i\om g_{11}(r_h)}~.
\ee
Moreover notice that the damping term comes with the linear $\om$ term, signaling ohmic stochastic dynamics of the particle in thermal theory.

Let us express all our results in terms of the temperature of the theory \eq{temperature}. The linear response is
\be\la{responset}
\chi(\om)=\frac{2\pi \a'}{-i\om} \prt{\frac{4\pi}{a_f}}^{2\n-1} ~ T^{1-2\n}~,
\ee
while the diffusion constant reads \eq{diffusion1}
\be\la{diffusiont}
D=2\pi\a'  \prt{\frac{4\pi}{a_f}}^{2\n-1}~ T^{2\prt{1-\n}}~.
\ee
The power depends on the order of the Bessel function defined in \eq{def1} related with an algebraic relation on the scalings of the background metric elements. The constant may increase with the temperature for $\nu<1$, or decrease  for $\n>1$. The crossover value $\n=1$ is the same with the one obtained in two point function \eq{twopointf} and \eq{crossover} of the zero temperature quantum fluctuations.

\subsection{Response Function, Fluctuation-Dissipation and The Nature of Thermal Noise }

Working as in the previous section and taking into account higher $\om$ terms, the response function is written as
\be
\chi(\om)=2\pi \a'\prt{\frac{i}{\g \om}-\frac{m}{\g^2}+\cO(\om)}~,
\ee
where
\be\la{thermalcor}
\g= r_h^{a_1}~,\qquad m= r_h^{2 a_1} \prt{-c_0+ \frac{r_b^{-2\k\n}}{2\k\n}}+ m_0~,
\ee
are the damping coefficient already produced in \eq{responsemet} and the inertial mass is thermally corrected with respect to the zero temperature $m_0$ obtained in \eq{massin}. The thermal contribution to the mass depend on the temperature as expected, on the boundary cut-off of the string and the integration constant $c_0$.  This is natural since the length of the fluctuating string can be traded with the energy needed to create the string.

The response function is independent of the constant $c_1$, in contrast to the correlator of the fluctuations, so we need to determine the constant. The mode expansion has divergences which can be canceled by requiring discreteness of frequencies, in agreement with the nature of boundary conditions.
Having density of states $\D\om^{-1}=-\log{\e}/(4\pi^2 T)$ the mode expansion reads
\be
x(t,r)= \sqrt{\frac{-\log\e}{4\pi^2 T}}\int_0^\infty d\om \prt{a_\om e^{-i \om t}+a^\dagger_\om e^{i\om t}}~.
\ee
We use the Klein-Gordon type product \eq{inner1} and similar discussion for the quantization condition of the zero temperature analysis in section \ref{section:QFTT}, to determine the unknown constant. The computation is quite involved with long expressions so we refrain from presenting here the full detail.

The main contributions to the integral from the inner product conditions \eq{uconditions}, come from the
horizon and the boundary area. Therefore, one needs to split the integral \eq{inner1} to two integrals
performing the integration and ignoring the intermediate $r$ contributions. The string solutions that contribute to the $c_1$ value are the $h_{Ah}(r)$ \eq{aa1} and $h_{Cb}(r)$ \eq{hcresult}. Schematically the constant $c_1$ is given by
\be
c_1^2=\frac{1}{2 i \om}\prt{\int_r^{r_b} \frac{r^{a_1-\frac{1}{2}\prt{a_0+a_u}}}{f(r)} h_{Ah}(r)+\int_{r_h+\e}^r \frac{r^{a_1-\frac{1}{2}\prt{a_0+a_u}}}{f(r)} h_{Cb}(r)}~,
\ee
and taking the leading contributions. Therefore our method applies at certain limits to gravity solutions, that have boundary and IR behavior belonging to the class of metric \eq{polymetric} including RG flow solutions.

By making the natural assumption that in the canonical ensemble the excited modes for thermal distribution are of Bose-Einstein type we get
\be\la{fdthermal}
\vev{X_1(\om) X_1(0)}\sim \frac{1}{1-e^{-\frac{\om}{T}}} \frac{\log{\e}}{r_h^\k} \frac{ r_h^{\k-a_1}}{ \om \log{\e}}\sim \frac{1}{1-e^{-\frac{\om}{T}}} \frac{1}{\om r_h^{a_1}}~,
\ee
where in the normalization factor we have traded the temperature with the horizon value using \eq{temperature} and in the intermediate equation we show the different contributions.
Therefore, the fluctuation-dissipation theorem \eq{theoremfd} in the thermal case holds for the whole class of theories considered here and at certain limits of RG flows with UV and IR asymptotics belonging to the class of theories \eq{polymetric}.

The correlator of the random force $\xi$, can be computed now in a straight-forward way
\be\la{rforce}
\vev{\xi_\om ~ \xi_{-\om}}:=\frac{\vev{X_1(\om) X_1(-\om)}}{\chi(\om)\chi(-\om)}\sim T^{2\n}~,
\ee
which shows the independency of the corresponding real time correlator on the frequency $\om$ related to the fact that the random force is a white noise and the simple dependence on the order on the Bessel function.

\section{Quantum Fluctuations of a Dp-brane in a $d+1$-spacetime} \la{sec:dpfluc1}

In this section,   we consider the quantum fluctuations on Dp-branes.
We show that for rigid branes there is a direct way to map the dynamics of their fluctuations  to the string fluctuations. After the mapping is done the two point-function of the brane fluctuations, the response function and the diffusion constant can be
read directly from our results in previous sections.

Let us consider the Dp brane in the d+1 space-time \eq{gen1}. We parametrize the brane using the radial gauge: $x_0=\t,~r=\s_1:=\s, x_2=\s_2,\ldots, x_p=\s_p, x_j=x_j(\t,\s_i)$, where 
 $j$ runs on the transverse dimensions of the brane $j=1, p+1,\ldots,d$. The reason of such a parametrization will become soon obvious. The Dirac-Born-Infeld(DBI) action reads
\be
S=T_p\int dx^{p+1}\sqrt{-\prt{g_{00}+D_t X}\prt{g_{uu}+D_1 X}\prt{g_{22}+D_2 X}\ldots\prt{g_{p p }+D_{p}X}} ~,
\ee
where $D_i X := g_{11}\pp_i X_{1}^2+g_{p+1}\pp_{p+1} X_{p+1}^2+\ldots+g_{dd}\pp_i X_d^2$, the operator acting on the transverse to the brane coordinates. Like the string, a solution to the DBI action is a localized brane on the transverse space. By considering the fluctuations along the spatial transverse direction $x_1$ we obtain
\be
S_{\mbox{DBI},2}=\frac{T_p }{2} \int d\t d\s_i \prt{-\frac{g_{11} \sqrt{-g_{00}}\sqrt{g_{22}\ldots g_{p p}}}{\sqrt{g_{rr}}}\d x_1'^2+ \frac{g_{11}\sqrt{g_{rr}}\sqrt{g_{22}\ldots g_{pp}}}{\sqrt{-g_{00}}}\d \dx_1^2}~.
\ee
The equation of motion for a solution of the form $\d x_1 (\t,\s)=e^{-i\o \t} h_\om(r)$ reads
\be \la{eqbmodes1}
\pp_r\prt{-\frac{g_{11} \sqrt{-g_{00}}\sqrt{g_{22}\ldots g_{pp}}}{\sqrt{g_{rr}}}h_\o(r)'}- \frac{g_{11}\sqrt{g_{rr}}\sqrt{g_{22}\ldots g_{p p }}}{\sqrt{-g_{00}}}\o^2 h_\om(r)=0~,
\ee
which can be solved to analyze the brane modes. Alternatively, we show we can map the brane fluctuations to the string fluctuations in a one-to-one way by relating the relevant equations. This is doable, due to the fact that the brane is 'rigid'. Similar mappings with extra conditions needed, have been observed between Dp-branes related to $k$-strings, high representations Wilson loops and fundamental strings \cite{Giataganas:2015ksa}.

The equations \eq{eqmodes1} and \eq{eqbmodes1} suggest the following substitution to the equation \eq{eqmodes1}
\be\la{subs1}
g_{11}  \rightarrow   g_{11}\sqrt{g_{22} g_{33}\ldots g_{dd} }~,
\ee
to obtain the brane fluctuations. Using the spacetime scaling \eq{polymetric} with the substitution
\be\la{eqa1}
a_1\rightarrow \tilde{a}_1=a_1+\frac{1}{2}\prt{a_2+\ldots+a_p}~,
\ee
in the string fluctuation equation \eq{eqmodes2} we obtain the brane fluctuation equation. Therefore the analysis does not have to be repeated and we simply list the following results.
Let us define the quantities from \eq{def1} using \eq{eqa1} as
\be\la{defb1}
\tilde{\nu}:=\frac{a_0+2 \ta_1+a_u-2}{2\prt{a_0+a_u-2}}~,\qquad \trr:=\frac{2 r^{\frac{1}{2}\prt{2-\a_0-\a_u}}}{a_0+a_u-2}=\frac{r^{-\k}}{\k}~,\qquad \k:= \frac{a_0+a_u}{2}-1~.
\ee
By making the sensible assumption that at late times the dominant contribution to the two-point function comes from the low frequency limit of \eq{twopointen}, we obtain
 \be\la{twopointbf}
\vev{X_1(t) X_1(0)}\sim
\begin{cases}
E^\frac{4\k\prt{1-\tilde{\n}}}{ \prt{a_0-\k}} ~|t|^{3-2\tilde{\n}}~,\quad \rm{when}\quad \tilde{\n} \ge 1~, \\
E^0 ~|t|^{ 2\tilde{\n}-1}~,\quad \rm{when}\quad \tilde{ \n } \le 1~.
\end{cases}
\ee
The parameter space is shifted by the modification of the extra brane dimensions by $\tilde{\n} \ge 1$ or $\tilde{\n}<0$. The crossover occurs at
\be\la{crossoverb}
\tilde{\n}=1\Leftrightarrow 2\a_1=a_u+a_0-\prt{a_2+\ldots+a_p}-2~,
\ee
which when saturated the two-point function grows linearly with $t$. In the positive $\tilde{\nu}$ region this is the maximum rate of growth of the two point function. The minimum rate of growth can be realized when $\tilde{\n}=3/2$ and $\tilde{\n}=1/2$ which result to a logarithmic behavior of the two point function. The response function then for the quantum brane fluctuations reads
\be\la{responseb1}
\chi(\omega) =\frac{2\pi\a'}{\om} r_b^{-a_1-\frac{1}{2}\prt{a_2+\ldots+a_p}} \frac{H_{\tilde{\n} }(\om \trr_b)}{H_{\tilde{\n}-1}(\om \trr_b)}~.
\ee
Let us now look at the low-frequency expansion, as defined in \eq{smallo}, of the response function. It takes the form
\be
\chi(\om)\sim -c_1\prtt{ m\prt{i\om}^2 +\g \prt{-i\omega}^{2\tilde{\n}}+\ldots}^{-1}~,
\ee
where $m$ is the inertial mass and $\g$ is the self-energy of the state
\be
m=\frac{r_b^{2\k\prt{\tilde{\n}-1}}} {2 \k \prt{\tilde{\n}-1}}~,\qquad \g=\frac{1-i\tan\prtt{\prt{\tilde{\n}-\frac{1}{2}}\pi}}{\prt{ \prt{2 i \k}^{2\tilde{\n}-1}\G\prt{\tilde{\nu}}^2}} \pi ~.
\ee
For $\tilde{\n}<1$ the self-energy dominates over the inertial mass at low frequencies.  We also notice that there is an allowed region for the scaling parameters where the mass becomes negative or diverges $2a_1\le a_0+a_u-\prt{a_2+\ldots+a_p}-2$. Moreover, the fluctuation-dissipation theorem for a brane propagating in generic spacetime at zero temperature is satisfied.

Going to the finite temperature the response function for the rigid brane can be read from \eq{chifint}
\be
\chi(\om)=\frac{2\pi \a'}{-i\om r_h^{a_1+\frac{1}{2}\prt{a_2+\ldots+a_p}}}~.
\ee
In the low frequency limit, following the relevant string discussion, we propose that for any arbitrary background it takes the following form
\be
\chi(\om)=\frac{2\pi \a'}{-i\om g_{11}(r_h)\sqrt{g_{22}(r_h)\ldots g_{pp}(r_h)}}~.
\ee
The diffusion constant  in terms of the temperature of the theory \eq{temperature} reads
\be\la{diffusionbf}
D=2\pi\a'  \prt{\frac{4\pi}{a_f}}^{2\tilde{\n}-1}~ T^{2\prt{1-\tilde{\n}}}~
\ee
The constant increases with the temperature for $\tilde{\nu}<1$, and decreases for $\tilde{\n}>1$, while the crossover value $\tilde{\n}=1$ is the same with the one obtained in two point function \eq{twopointbf} and \eq{crossoverb} of the zero temperature brane quantum fluctuations.  Moreover, the fluctuation-dissipation theorem at finite temperature is satisfied.

\section{Application To Various Gravity Dual Theories} \la{sec:appl}

Let us demonstrate how our findings apply to particular gravity dual field theories belonging in the class of backgrounds of \eq{gen1} and \eq{polymetric}.

\subsection{Anisotropic Quantum Critical Points}
\subsubsection{Lifshitz-like with Fixed Scaling Coefficient}
Anisotropic IIB supergravity solutions dual to Lifshitz-like fixed points with anisotropic scale invariance were found in \cite{Azeyanagi:2009pr}. In string frame they read
\be
ds^2_s=\ti{R}_s^2\left[r^{\frac{7}{3}}\prt{-f(r)dt^2+dx^2
+dy^2}+r^{\ff{5}{3}}dw^2+\ff{dr^2}{r^{\frac{5}{3}}f(r)}\right]
+R_s^2r^{\frac{1}{3}}ds_{X_5}^2, \la{aze1}
\ee
where $R_s^2=\ff{12}{11}\ti{R}_s^2 $. And the dilaton and the blackening factor read
\be
e^{\phi}=r^{\ff{2}{3}}e^{\phi_0}, \qquad f(r)=1-\prt{\frac{r_h}{r}}^{\ff{11}{3}} \la{aze2}
\ee
where $e^{\phi_0}=\frac{\sqrt{22}}{3\a}$. The scaling is fixed and equal $z=3/2$ due to fixed form of the axion-dilaton coupling form in the IIB supergravity \cite{Giataganas:2017koz}. The finite temperature RG flows with AdS boundary flowing to the above Lifshitz-like space have been found in \cite{Mateos:2011ix} and the analysis of the heavy quark observables revealed very interesting properties, including the diffusion \cite{Giataganas:2012zy,Chernicoff:2012iq} and the Langevin coefficients \cite{Giataganas:2013hwa,Giataganas:2013zaa}; where a review can be found in \cite{Giataganas:2013lga}. The above class of solutions has been shown recently that belongs to a generalized Lifshitz-like hyperscaling violation background with arbitrary scalings described below.

\subsubsection{Arbitrary Hyperscaling Violation Lifshitz-like Geometries}

The generalized RG  anisotropic flows were found in \cite{Giataganas:2017koz}. Here we consider the generalized  Einstein-Axion-Dilaton action  with a  potential for the dilaton and an arbitrary coupling  between the axion and the dilaton to find the gravity solution in the string frame. The action reads:
 \be\label{actiongiata}
S=\frac{1}{2\kappa^2}\int d^{5}x\,\sqrt{-g}\left[R-\frac{1}{2}(\partial \phi )^2+ V(\phi ) -\frac{1}{2} Z(\phi ) (\partial \chi )^2\right].
\ee
with a  choice of functions
\be
V(\phi)=6 e^{\sigma\phi},\qquad Z(\phi)=e^{2\gamma\phi}.
\ee
The solution to the system of equations generated by \eq{actiongiata} is a Lifshitz-like anisotropic hyperscaling violation  metric which exhibits the arbitrary critical exponent $z$ and a hyperscaling violation exponent $\th$ related to the constants the $\s$ and $\g$. The geometry may also accommodate a black hole reading as
\bea\la{hyscam2}
 ds_s^2=a^2 C_R e^{\frac{\phi(r)}{2}} r^{-\frac{2\th}{dz}} \prt{-r^{2} \prt{f(r)dt^2+dx_i^2}+C_Z r^{\frac{2}{z}} dx_3^2+\frac{dr^2}{f(r)a^2 r^2}} ~,
\eea
where
\be\la{fdil}
 f(r)=1-\prt{\frac{r_h}{r}}^{d+\prt{1-\th}/z}~,\qquad e^{\frac{\phi(r)}{2}}=r^\frac{\sqrt{\th^2+3 z\prt{1-\th}-3}}{\sqrt{6}z}~.
\ee
The constants $C_R,~C_Z$ and the scalings $\th,~z$ depend on the parameters $\g$ and $\s$ of the action with analytic algebraic relations. The Hawking temperature of the theory is
\be
T=\frac{|d+(1-\th)/z|}{4 \pi r_h^z} ~.
\ee
We choose the spatial dimensions as $d=3$. The solution \eq{hyscam2} and \eq{fdil} becomes of IIB supergravity for $\g=1,~\s=0$ corresponding to $z=3/2,~\th=0$ reproducing the geometry \eq{aze1}, \eq{aze2}. The scaling constants $(z,\th)$ and consequently the parameters $(\s,\g)$ are constrained by the null energy conditions ensuring attractive gravity
\bea  \nn
&&\th^2+3z\prt{1-\th} -3\ge 0~,\\\la{ineq1}
&&\prt{z-1}\prt{1+3 z -\th}\ge 0~
\eea
and the thermodynamic stability conditions and more particularly the positivity of the specific heat reads
\be\la{sph}
 2+\frac{1-\th}{z}\ge 0 ~.
\ee
Notice that the condition from the null energy condition \eq{ineq1}, ensures that the dilaton \eq{fdil} and our metric has no imaginary part. At the zero temperature limit the last condition is not necessary, however zero temperature theories with scalings that do not satisfy this inequality, may have entanglement entropy that scales faster than the area of the entangled region considered, signaling instabilities.

\begin{figure}
\begin{minipage}[ht]{0.5\textwidth}
\begin{flushleft}
\centerline{\includegraphics[width=70mm]{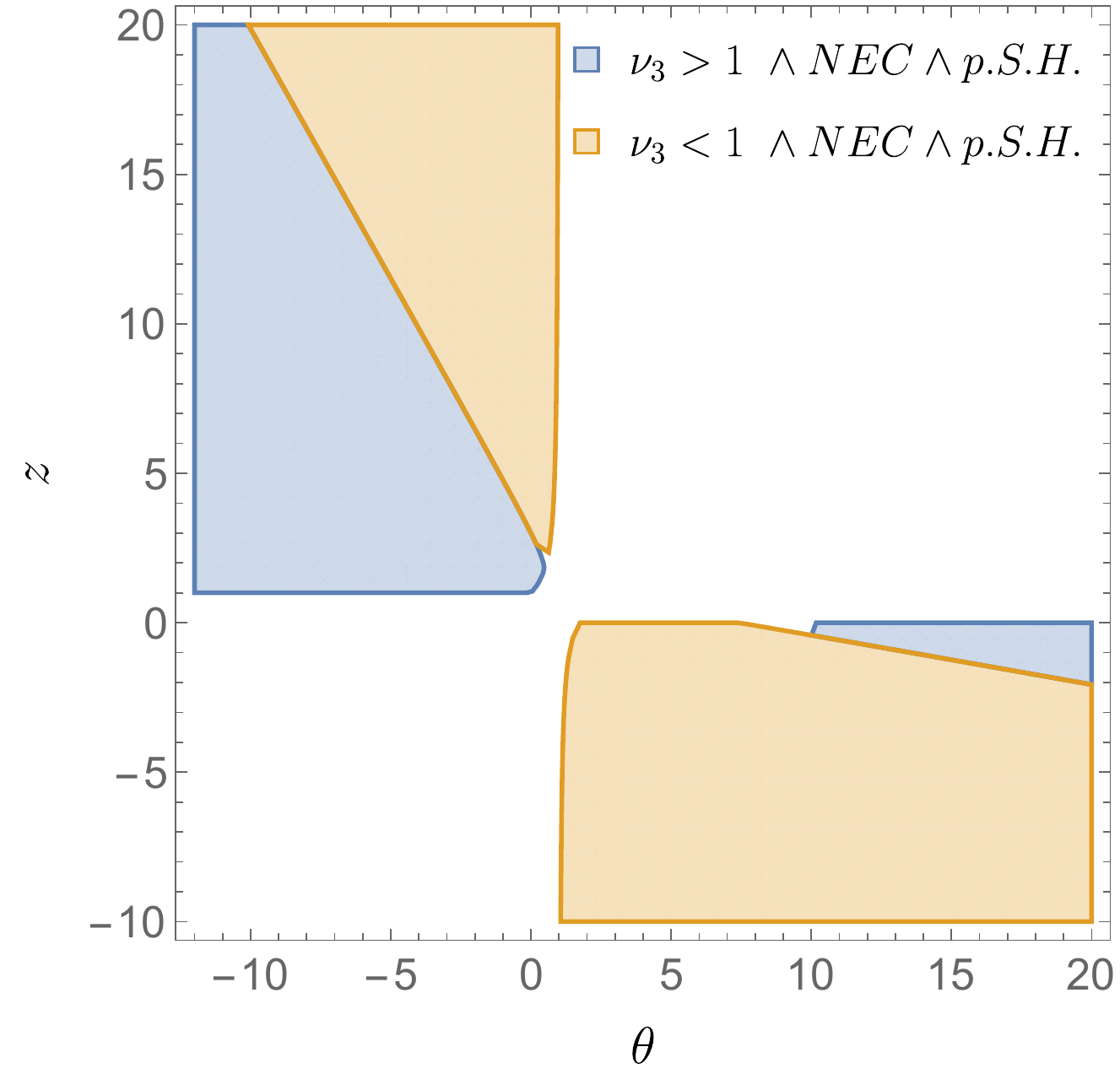}}
\caption{\small{Sample of the physical theories satisfying in addition the inequality \eq{sph} for fluctuations along the $x_3$ direction, which belong to the upper and lower branches of two-point function \eq{twopoint1}.}}
\label{fig:range1}
\end{flushleft}
\end{minipage}
\hspace{0.3cm}
\begin{minipage}[ht]{0.5\textwidth}
\begin{flushleft}
\centerline{\includegraphics[width=70mm ]{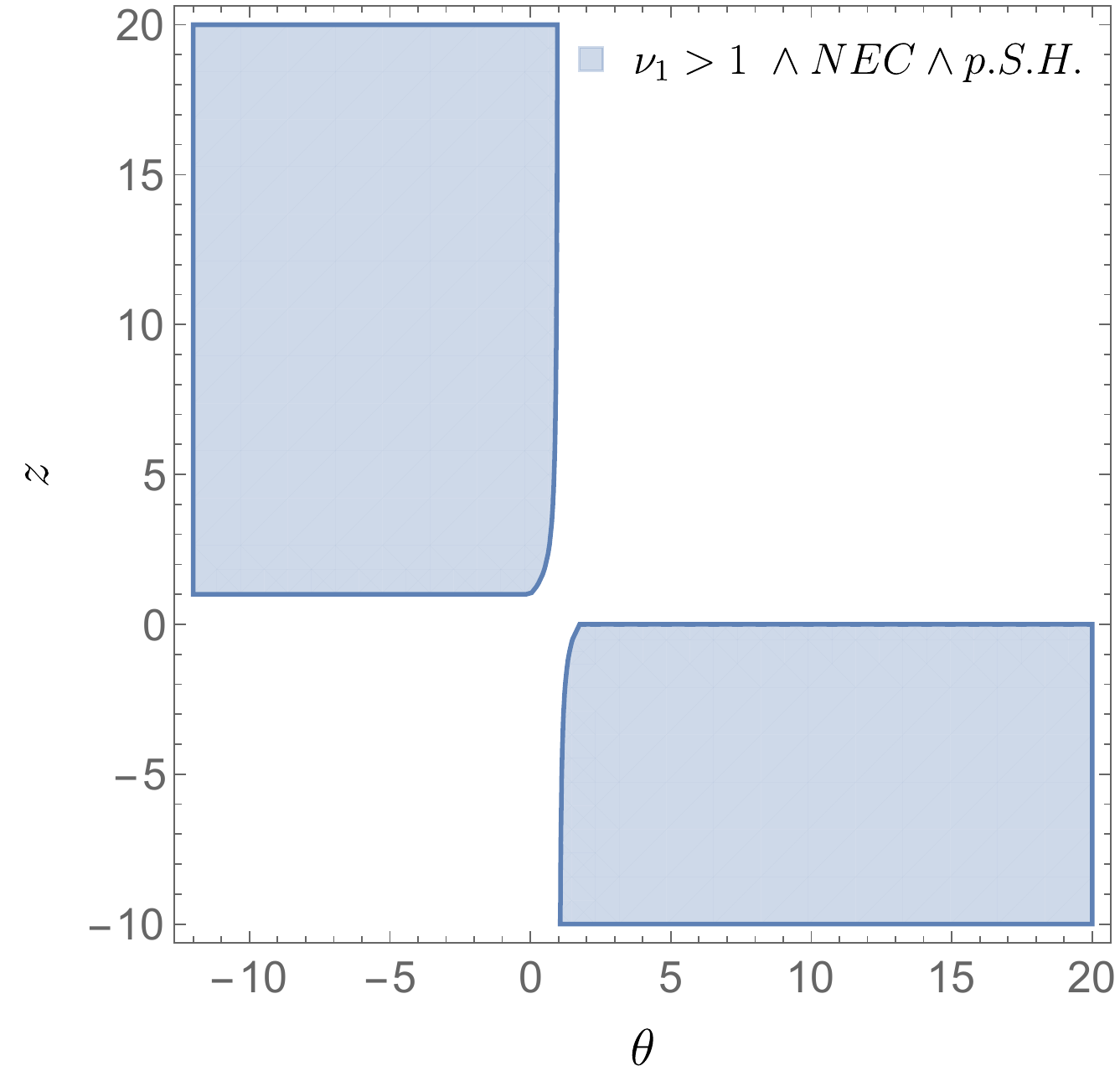}}
\caption{\small{Sample of the stable and  physical theories   for fluctuations on the transverse plane \eq{twopoint2}.}}
\label{fig:range2}\vspace{.7cm}
\end{flushleft}
\end{minipage}
\end{figure}

Let us consider fluctuations along the anisotropic direction $x_3$ focusing on the scalings of the metric elements. The parameters $a_i$ can be read easily from the metric \eq{hyscam2} and the equations \eq{fdil} to give for the order of the Bessel function
\be
\nu_3=\frac{12+6z-4\th +\sqrt{6}\sqrt{3z\prt{1-\th}-3+\th^2}}{12 z}~.
\ee
The crossover region for the two-point function and the response function is for $\n=1$. The two-point function \eq{twopointf} can be defined in both regions as shown in the Figure \ref{fig:range1}, to give \eq{twopointf},
 \be\la{twopoint1}
\vev{X_3(t) X_3(0)}\sim
\begin{cases}
E^{2\frac{12-6z-4\th +\sqrt{6}\sqrt{3z\prt{1-\th}-3+\th^2}}{-6z+4\th-\sqrt{6}\sqrt{3z\prt{1-\th}-3+\th^2}}} ~|t|^{3-2\n_3}~,\quad \rm{when}\quad \n_3 \ge 1 ~,~ Figure ~\ref{fig:range1}, \\
E^0 ~|t|^{ 2\n_3-1}~,\quad \rm{when}\quad  \n_3 \le 1~,~ Figure~ \ref{fig:range1}~.
\end{cases}
\ee
The rest of the observables, can be read directly from the main text of our paper since we have determined the order of the Bessel function $\nu$. For example the diffusion constant in the finite region area is given by \eq{diffusiont}
\be\la{diffusion3}
D_3=2\pi\a'  \prt{\frac{4\pi}{d+\prt{1-\th}/z}}^{2\n_3-1}~ T^{2\prt{1-\n_3}}~.
\ee

Fluctuations along the transverse plane $x_1-x_2$ lead to different behavior.  The order of the Bessel function is
\be
\nu_1=\frac{18z-4\th +\sqrt{6}\sqrt{3z\prt{1-\th}-3+\th^2}}{12 z}~.~
\ee
and all the observables along this direction scale differently than above. The crossover region is again for $\n_1=1$. However in this case, the NEC and the inequality \eq{sph}, enforce all the physical and stable theories to live in one region (Figure \ref{fig:range2}); for $\nu<1$ there is no theory with the desirable properties, so we have
\be\la{twopoint2}
\vev{X_1(t) X_1(0)}\sim E^{-2} ~|t|^{3-2\n_1}~,\quad \rm{when}\quad \n_1 \ge 1  ~,~ Figure~ \ref{fig:range2}~.
\ee
Notice that the dependence on the energy is fixed for any scaling parameter. Having specified the order $\nu_1$ we can derive all the theory observables. For example the diffusion along the transverse plane will now scale as $D_1=~ T^{2\prt{1-\n_1}}~$ comparing it to the \eq{diffusion3}. The self energy and the inertial mass receives thermal correction that depend on the direction of fluctuations, using  \eq{thermalcor} we get
\be
\g_1= r_h^{a_1}~,~ m_1= r_h^{2 a_1} \prt{-c_0+ \frac{r_b^{-2\k_1\n_1}}{2\k_1\n_1}}+ m_{0{_1}}~, ~ a_1:=2-\frac{2\th}{3 z}+\frac{\sqrt{3z\prt{1-\th}-3+\th^2}}{\sqrt{6}}~,
\ee
where $\k_1=a_1/\prt{2 \n_1-1}$ and $m_{0_{1}}$ is the inertial mass at zero temperature which depends on the direction of the fluctuation given by \eq{massin}. Moreover, we point out that the fluctuation-dissipation theorem is satisfied for anisotropic theories when considered along each direction.

Let us also briefly discuss the anisotropic IIB background \eq{aze1} which is obtained for $\th=0,z=3/2$ and by following the above notation gives the for anisotropic and transverse fluctuations
\be
\nu_3=\frac{4}{3}~,\qquad \nu_1=\frac{5}{3}~.
\ee
Having specified the properties of the string fluctuations in these backgrounds we can move on to read the Dp-brane ones. The shifted $\tilde{\n}$ parameter will now depend on the dimension of the brane and give modified equations compared to the string ones derived above. We refrain from giving the analytic results since the expressions are quite longer and are easy to be read from the string analysis above and the formulas of the main text. Instead we give only the order of the Bessel function for Dp-brane fluctuations in the IIB supergravity solution \eq{aze1}
\be
\tilde{\nu}_3=\frac{4}{3}+\frac{7 p}{12}~,\qquad \tilde{\nu}_{1,a}=\frac{3}{2}+\frac{7p}{12}~,\qquad \tilde{\nu}_{1,b}=\frac{5}{3}+\frac{7p}{12}~,
\ee
where for the transverse fluctuation for the $\tilde{\nu}_{1,a}$ we have considered one of the p-directions as the anisotropic $x_3$, while for the $\tilde{\nu}_{1,b}$ none of the $p$ directions are anisotropic. All the observables follow from the main text analysis, and will depend on the number of the brane directions.

\subsection{Quantum Critical Points}
In this section we present very briefly how some of the known results together with some new ones on quantum critical points are covered in our universal analysis under the scheme we provide.

Let us consider the hyperscaling violation metrics
\bea\la{hyscam33}
 ds_s^2=r^{-\frac{2\th}{d  }} \prt{-r^{2z} f(r)dt^2+r^2 dx_i^2+ \frac{dr^2}{f(r) r^2}} ~,
\eea
then the order of the Bessel function reads
\be
\nu=\frac{1}{2}+\frac{1}{z}-\frac{\th}{dz}
\ee
and the two point function follows
\be\la{twopoint3}
\vev{X_1(t) X_1(0)}\sim
\begin{cases}
E^{\frac{2\prt{-2d+dz+2\th}}{dz-2\th}} ~|t|^{2-\frac{2}{z}+\frac{2\th}{3z}}~,\quad \rm{when}\quad \n  \ge 1 ~, \\
E^0 ~|t|^{\frac{2}{z}-\frac{2 \th}{3z}}~,\quad \rm{when}\quad  \n  \le 1~,
\end{cases}
\ee
agreeing with the correlation obtained for Lifshitz critical points \cite{Tong:2012nf} and the hyperscaling violation critical points \cite{Edalati:2012tc}. The zero temperature result in this case, can be seen as a reformulation of \cite{Edalati:2012tc}, since with appropriate coordinate transformation, the hyperscaling violation metric which has two unknown scalings and can be related to the metric ansatz we have.

In finite temperature we can get in a straightforward way our observables. For example the self energy and the inertial mass receives thermal correction \eq{thermalcor}
\be
\g= r_h^{a_1}~,~ m_1= r_h^{2 a_1} \prt{-c_0+ \frac{r_b^{-2\k\n}}{2\k\n}}+ m_0~,~m_0=\frac{r_b^{2\k\prt{\n-1}}} {2 \k \prt{\n-1}}~,~ a_1:=2z-\frac{2\th}{3}~,~ k=\frac{a_1}{\prt{2 \n-1}}~,
\ee
where $\nu$ is given by \eq{twopoint3}, the same with the zero temperature analysis and $m_0$ is the zero temperature inertial mass. By making use of our main text analysis we can obtain  the rest of the observables discussed in our text. It is also straightforward from the Dp-brane analysis of the section \la{sec:dp} to compute the scalings and the shifted order of the Bessel function $\tilde{\n}$, to get in this case in a straightforward way the results of \cite{Yeh:2013mca} in Lifshitz space-time and new ones in the hyperscaling violation space-time.

The minimum rate of growth for the string two-point function can be realized when $\n=3/2$ and $\n=1/2$ which result to a logarithmic behavior of the two point function. The $\cN=4$ sYM dual to AdS gravity dual theory, give $\n=3/2$ and exhibiting the logarithmic behavior \cite{deBoer:2008gu}.

\section{Brief List of Results and Summary Remarks} \la{sec:conc}

Due to the technical nature of this work, and the direct information that
the reader can obtain from the derived formulas, let  us first give a brief
guide to the main equations of the paper and present in the next subsection our final remarks. An analytic discussion on our results and the impact of our work has been presented already in section \ref{sec:impact}.

\subsection{Brief List of the Main Equations of Our Work:}

\textbf{Holographic Theories:}  We provide a unified scheme for the study of quantum and thermal fluctuations in a wide class of holographic theories given by \eq{gen1} and \eq{polymetric}, directly applicable to a large class of theories. The equation of motion we to solve for string fluctuations in generic spacetime is given by \eq{flucgen}.\newline
\textbf{Quantum fluctuations:}  The quantum fluctuation analysis turns out to be described by Bessel functions \eq{solmodes2} of order $\n$, which depends on the arbitrary scalings of the metric elements  \eq{def1}.  The two point function \eq{twopointf} of the zero temperature fluctuations exhibits two different late time behaviors depending purely on the value of the order of Bessel function, and there is always a crossover region for $\n=1$.  The response function is computed in the class of holographic theories and the fluctuation-dissipation theorem is verified from the bulk perspective to be always satisfied \eq{flucdis}.  The inertial mass and the self energy of the particle are read from the low frequency limit of the response function \eq{massin}. Between them there is a competition for dominance in the response, which depends exclusively on the order of Bessel function, while the critical value is again $\n=1$. One of the main focuses of this section, is
that our formulation shows that    our stochastic observables depend mainly on the order of the Bessel function.\newline
\textbf{Thermal fluctuations:} We study the boundary particle fluctuation by bringing their ruling equation to a Schr\"{o}dinger-like form \eq{schr}-\eq{yeq}; and  employing a monodromy patching method \eq{eq31}-\eq{comb3153} to solve it.  The response function of the thermal fluctuations is inversely proportional to the frequency and the value of the metric element at the horizon along the direction that the fluctuation occurs \eq{responsemet}; we suggest that this relation is true even for other theories do not belong in the class considered here. The scaling of the response function with respect to the temperature depends solely on the order of Bessel function and is $T^{1-2 \n}$ \eq{responset}.  While the diffusion constant   depends on the temperature as $T^{2\prt{1-\n}}$ \eq{diffusiont} and the inertial mass receives a thermal correction \eq{thermalcor}. The fluctuation-dissipation theorem holds \eq{fdthermal} for the whole class of the backgrounds, while the random force is white noise with the self correlator scaled with the temperature as $T^{ 2 \n}$ \eq{rforce}.\newline
\textbf{Dp-brane Fluctuations:} We study the rigid Dp-brane fluctuations in a $d+1$ dimensional space and prove that there is one-to-one map to the string fluctuations \eq{subs1}.  Therefore we are able to determine in a straightforward way the Dp-brane observables: two-point function, response function, inertial mass, self energy, thermal response function and the diffusion constant: \eq{twopointbf}-\eq{diffusionbf}. All the Dp-brane observables depend on the shifted $\tilde{\n}$ parameter \eq{defb1}, in analogy with the string results.

\subsection{Final Remarks}

Motivated by the wide range of applicability of the fluctuation and dissipation phenomena,  we provide a universal study scheme for quantum and thermal fluctuations, the dissipation of the energy and the corresponding Brownian motion of the particle, for a wide class of strongly coupled field theories.
A key-point of our work is that our analysis applies and covers several different dual field theories in a universal way. In the theory of fluctuations the total information of the different backgrounds is incorporated in a compact way, in the order of the Bessel solutions of the string mode equations and their argument. The analysis is carried out in full generality for the  order of the Bessel function solutions. The different backgrounds are associated with different values of the order $\nu$ with a simple algebraic relation. Therefore, one can extract straightforwardly from our methods,  as an application to special cases,  the study of Brownian motion and the corresponding formulas obtained for example in AdS, Lifshitz and hyperscaling violation backgrounds \cite{deBoer:2008gu,Son:2009vu,Tong:2012nf,Edalati:2012tc,Yeh:2013mca}, while our results cover and apply to a wide class of other theories.

It is worthy to point out that such universal treatments have been observed before in holography while studying completely different topics. For example in \cite{Giataganas:2014hma}, working with a similar wide range class of theories, the non-integrability conditions usually accompanied with the presence of chaos were obtained. The analysis there boils down to a simple algebraic condition for the order of the Bessel functions, related like here to the arbitrary scaling parameters of the metric. Together with our current work these independent studies strongly suggest that holographic metrics with polynomial arbitrary scalings may admit certain unified scheme analyses in holography.

Let us also comment on a further potential extension of our studies on the quantum phase transitions, which show abrupt changes in the ground state properties of a quantum many-particle system
when a non-thermal control parameter is varied. In particular, the impurity quantum phase transitions are an interesting class of quantum phase transitions in which, the systems can be composed of
a single quantum spin coupled to an infinite fermionic or bosonic bath. The prototypical system consists of a two-level state coupled to a single dissipative bath of harmonic oscillators, also called the spin-boson mode \cite{twostate1,twostate2,qpt22,qpt11}.  It will be of great interest to reexamine the nature of quantum phase transitions when the bath degrees of freedom with the general bath spectral density are strongly coupled using the holographic approach.

\subsection*{Acknowledgments}

We would like to thank J. Pedraza and K. Zoubos, for  useful conversations and correspondence. The work of DG is supported by  the National Center of Theoretical Science (NCTS) and the grants  101-2112-M-007-021-MY3 and 104-2112-M-007 -001 -MY3 of the Ministry of Science and Technology of Taiwan (MOST). The work of DSL and CPY is supported in part by the MOST. We are also grateful to the MOST and NDHU for supporting a visit by DG to the NDHU during the completion of the project.

\begin{appendices}

\section{Bessel Function Properties} \label{appendix:bes}

We present briefly some of the key properties of the Bessel functions that have allowed us to obtain the
solutions of the equations of motion in full generality.
The first and second kind Bessel functions are related as
\bea
&&J_\n(r)=\frac{1}{\sin \pi \n} Y_{-\n}(r)-\cot \pi \n Y_\n(r)~,\\
&&Y_\n(r)=\frac{1}{\sin \pi \n}\prt{\cos\pi\n J_{\n}(r)- J_{-\n}(r)}~.
\eea
The orthogonality and normalization conditions are
\bea
&&\int dr ~r J_\n(\omega r) Y_\n(\tilde{\omega} r)=0,\\
&& \int dr~ r J_\n(\omega r) J_\n(\tilde{\omega} r)=\int dr r Y_\n(\omega r) Y_\n(\tilde{\omega} r)=\frac{1}{\omega} \d\prt{\omega-\tilde{\omega}}~.
\eea
The following identities involve derivatives hold for both first and second kind Bessel functions $J_\n,Y_\n$:
\bea
&&J_\n'(r)=J_{\n-1}(r)-\frac{\n}{r} J_{\n}(r)~,\\
&&J_\n'(r)=\frac{1}{2}\prt{J_{\n-1}(r)-J_{\n+1}(r)}~.
\eea
The identity that relate the Bessel functions of different kind and consecutive neighbors is
\be
J_\n(r) Y_{\n-1}(r)-J_{\n-1}(r) Y_{\n}(r)=\frac{2}{\pi r}~,
\ee
while  for the first Hankel function reads
\be
H'(r)=H_{\n-1}(r)-\frac{\n}{r} H_{\n}(r)~.
\ee

\section{Coordinate transformation to bring the Fluctuation ODE to Schr\"{o}dinger-like form} \label{app:coor}

The transformation is of general interest for the application of the monodromy patching method on the solution of the ordinary differential equation. Let us consider the  differential equation
\be\la{ode1}
\frac{\pp }{\pp r}\prt{r^\a f(r) \frac{\pp x(r)}{\pp r}}+\frac{\om^2}{r^\b f(r)}x(r)=0~,
\ee
where $f(r)$ is an arbitrary function, in our case the blackening factor, while $\a$ and $\b$ are constant powers. We  bring the differential equation \eq{ode1}, which of the same form of the \eq{eqmodes22}, to
\be\la{ode2}
\frac{\pp^2 y(r)}{\pp r_\star^2}+ \prt{\om^2-V(r )}y=0~.
\ee
where
\bea
&&y=x r^\frac{\a-\b}{4}~,\qquad  \frac{\pp r_\star }{\pp r}=\frac{1}{f(r) r^\frac{\a+\b}{2}}~,\\
&& V(r)=\frac{\b-\a}{4} f(r) r^{\a+\b-2}\prt{\prt{\frac{3\a+\b}{4}-1}f(r)+r f'(r)}~.
\eea
Using the transformation we can study the solution on overlapping regions according to the values of the ratio $V(r)/\om$, to find the approximate solution.


\section{Review of Generic Study of the Trailing String  }\la{app:trailing}

We briefly present some of the generic results of \cite{Giataganas:2013hwa,Giataganas:2013zaa} were the trailing string was studied in full generality.  The trailing string corresponds to a quark moving on the boundary along a spatial direction, choose it as $x_3$,  with a constant velocity. It is parametrized as
$t=\t, r=\sigma, x_3=v~t+\xi(r)$, and localized in the rest of dimensions.

The momentum flowing from the boundary to the bulk $\Pi^3_r$, is a constant of motion
\be
\xi'^2=-g_{rr} C^2\,\frac{g_{00}+g_{33}\,v^2} {g_{00}g_{33}\prt{C^{2}+g_{00}g_{33}}}~,\quad C:=2~\pi\,\alpha'\,\Pi^3_r ~.
\ee
There is a special value of $r_0$ ensuring that the imaginary part of the above ratio is zero.
It is found by solving the equation
\be
g_{00}(r_0)=-g_{33}(r_0)~v^2~,\label{pointa1}
\ee
and becomes equal to the black hole horizon for a static particle, $v=0$. The corresponding drag force is
\be
F_{drag,x_3}=-\frac{1}{2\pi\alpha'}\frac{\sqrt{-g_{00}(r_0)~g_{33}(r_0)}} {2\pi}=-\frac{v~ g_{33}(r_0)}{2\pi\alpha'}~,
\ee
and the friction coefficient can be found as
\bea
F_{drag}=\frac{dp}{dt}=-\eta_D p,\qquad \eta_D=\frac{g_{33}(r_0)}{2\pi\alpha' M_q \gamma_v}~,\label{etaD}
\eea
where $m_q$ is the mass of the particle, $p=m_q\,v\gamma_v$ and $\gamma_v:=\left(1-v^2\right)^{-1/2}$. The moving particle feels a temperature different than the one that the heat bath produces equal to
\bea\label{wst11}
 T_{ws}^2=\frac{1}{16\pi^2}
\Bigg|\frac{1}{g_{00} g_{rr}}\prt{g_{00}\,g_{33}}' \prt{\frac{g_{00}}{g_{33}}}'\Bigg|~\bigg|_{r=r_0}~,
\eea
where the static particle feels the heat bath temperature $T_{ws}=T$.

By considering the fluctuations of the classical trailing string solutions we can compute the Langevin coefficients.  The heavy particle moving with a constant velocity $v$ in the heat bath undergoes a Brownian motion with an effective equation of motion
\be
\frac{dp_i}{dt}=-\eta_{D\,ij}\, p^j+\xi_i\prt{t}~,\label{langp1}
\ee
where $\xi_i(t)$ is the random force generated by the medium.  The force distribution is related to the two-point correlators along the longitudinal and transverse to the direction of motion $\k_a=\prt{\k_L,\k_T}$:
\be
\vev{\xi_a\prt{t}\xi_a\prt{t'}}=\k_a\d\prt{t-t'}~.
\ee
The diffusion coefficients are given by
\be
\k_{a}= -2~T_{ws}\,\lim_{\omega\rightarrow 0}~\frac{{\rm Im}G^a_{R}(\omega)}{\omega}~, \la{diff11}
\ee
where $G_R$ is the anti-symmetrized retarded correlator. The generic formulas for the transverse and longitudinal string fluctuations and therefore the Langevin coefficients can be expressed in the background metric elements as   \cite{Giataganas:2013hwa}
\bea\la{langk1}
\k_T=\frac{1}{\pi\alpha'}~g_{kk}\bigg|_{r=r_0} T_{ws}~,\qquad~
\k_L=\frac{1}{\pi\alpha'}\,\frac{\left(g_{00}g_{33}\right)'} {g_{33}\,\left(\frac{g_{00}}{g_{33}}\right)'}\Bigg|_{r=r_0} T_{ws}~,
 \la{mpa222}
\eea
where the index $k$ denotes a particular transverse direction (e.g. $k=1$ for $x_1$) to that of the motion $x_3$ and no summation is taken. The $T_{ws}$ is given in terms of metric elements by \eq{wst11}. It follows that their ratio can be written in a compact universal form \cite{Giataganas:2013hwa,Giataganas:2013zaa}
\be\la{ratio11}
\frac{\k_L}{\k_T}=\frac{\prt{g_{00}g_{33}}'} {g_{kk}g_{33}\,\left(\frac{g_{00}}{g_{33}}\right)'}\Bigg|_{r=r_0}~,
\ee
and for isotropic theories it has been proved universally that $\k_L\ge \k_T$. For anisotropic theories the inequality can be reversed and the universality relation is violated $\k_L \lesseqgtr \k_T$.

The Einstein-like relations for motion of a particle in generic backgrounds have been also derived. Starting with the linearized Langevin equations \eq{langp1}
\be
\g_v^3\,m_q \d \ddot{x}_{L}=-\eta_{L}~\d \dot{x}_{L}+\xi_{L}~,
\qquad\gamma_v ~m_q \d \ddot{x}_{T}=-\eta_{T}\,\d \dot{x}_{T}+\xi_{T}~,
\label{ftt1}
\ee
the friction coefficients $\eta_{L,T}$ are related to the coefficients $\eta_{D}$ as
\bea
\eta_{T}=M_q ~\gamma_v~   \eta_{_{D,T}}~, \qquad \eta_{L}=M_q~\g_v^3\,\prt{\eta_{_{D,L}}+p~\frac{\rd \eta_{{_D,L}}}{\rd p}\bigg|_{p=M_q v \g_v}}~,\label{einn2}
\eea
and the coefficient $\k_a$ using \eq{diff11} can be written as
\be
\k_{a}=2\, T_{ws}\, \eta_{a}~.\la{kat}
\ee
The Einstein relations for  the diffusion and friction coefficients are satisfied by
\bea
\frac{\k_T}{\eta_{D,T}}=2M_q~ \g_v ~T_{ws}~.
\eea
For isotropic backgrounds this result is similar to the one obtained in \cite{HoyosBadajoz:2009pv,Gursoy:2010aa,Giataganas:2013zaa}. The zero velocity limits imply $r_0=r_h$, $T_{ws}=T$ and $\k_L=\k_T$ \eq{langk1}, in agreement with the formula derived in the text.

\end{appendices}


\bibliographystyle{JHEP}

\begin{thebibliography}{10}

\bibitem{brownianII}
M.~C. Wang and G.~E. Uhlenbeck, \emph{On the theory of the brownian motion ii},
  \href{https://doi.org/10.1103/RevModPhys.17.323}{\emph{Rev. Mod. Phys.}
  {\bfseries 17} (Apr, 1945) 323--342}.

\bibitem{brownian100}
P.~Hänggi and F.~Marchesoni, \emph{Introduction: 100years of brownian motion},
  \href{https://doi.org/10.1063/1.1895505}{\emph{Chaos: An Interdisciplinary
  Journal of Nonlinear Science} {\bfseries 15} (2005) 026101}.

\bibitem{brownianrel}
J.~Dunkel and P.~Hänggi, \emph{Relativistic brownian motion},
  {\emph{Physics
  Reports} {\bfseries 471} (2009) 1 -- 73}  [\href{https://arxiv.org/abs/0812.1996}{{\ttfamily
  0812.1996}}].

\bibitem{brownian111}
X.~Bian, C.~Kim and G.~E. Karniadakis, \emph{111 years of brownian motion},
  \href{https://doi.org/10.1039/C6SM01153E}{\emph{Soft Matter} {\bfseries 12}
  (2016) 6331--6346}.

\bibitem{Moore:2004tg}
G.~D. Moore and D.~Teaney, \emph{{How much do heavy quarks thermalize in a
  heavy ion collision?}},
  \href{https://doi.org/10.1103/PhysRevC.71.064904}{\emph{Phys. Rev.}
  {\bfseries C71} (2005) 064904},
  [\href{https://arxiv.org/abs/hep-ph/0412346}{{\ttfamily hep-ph/0412346}}].

\bibitem{vanHees:2004gq}
H.~van Hees and R.~Rapp, \emph{{Thermalization of heavy quarks in the
  quark-gluon plasma}},
  \href{https://doi.org/10.1103/PhysRevC.71.034907}{\emph{Phys. Rev.}
  {\bfseries C71} (2005) 034907},
  [\href{https://arxiv.org/abs/nucl-th/0412015}{{\ttfamily nucl-th/0412015}}].

\bibitem{Mustafa:2004dr}
M.~G. Mustafa, \emph{{Energy loss of charm quarks in the quark-gluon plasma:
  Collisional versus radiative}},
  \href{https://doi.org/10.1103/PhysRevC.72.014905}{\emph{Phys. Rev.}
  {\bfseries C72} (2005) 014905},
  [\href{https://arxiv.org/abs/hep-ph/0412402}{{\ttfamily hep-ph/0412402}}].

\bibitem{Romatschke:2004au}
P.~Romatschke and M.~Strickland, \emph{{Collisional energy loss of a heavy
  quark in an anisotropic quark-gluon plasma}},
  \href{https://doi.org/10.1103/PhysRevD.71.125008}{\emph{Phys. Rev.}
  {\bfseries D71} (2005) 125008},
  [\href{https://arxiv.org/abs/hep-ph/0408275}{{\ttfamily hep-ph/0408275}}].

\bibitem{landausp}
L.~D. Landau and E.~M. Lifshitz, \emph{{ Statistical Physics, Part I}},
  {\emph{Pergamon Press, Oxford} {\bfseries {Course of Theoretical Physics,
  Vol. 5, Third Edition }} (1980) }.

\bibitem{caldeira1983}
A.~Caldeira and A.~Leggett, \emph{Path integral approach to quantum brownian
  motion},
  {\emph{Physica
  A: Statistical Mechanics and its Applications} {\bfseries 121} (1983) 587 --
  616}.

\bibitem{Schwinger}
J.~Schwinger, \emph{Brownian motion of a quantum oscillator},
  \href{https://doi.org/10.1063/1.1703727}{\emph{Journal of Mathematical
  Physics} {\bfseries 2} (1961) 407--432}.

\bibitem{Feynman:1963fq}
R.~P. Feynman and F.~L. Vernon, Jr., \emph{{The Theory of a general quantum
  system interacting with a linear dissipative system}},
  \href{https://doi.org/10.1016/0003-4916(63)90068-X}{\emph{Annals Phys.}
  {\bfseries 24} (1963) 118--173}.

\bibitem{Grabert:1988yt}
H.~Grabert, P.~Schramm and G.~L. Ingold, \emph{{Quantum Brownian motion: The
  Functional inegral approach}},
  \href{https://doi.org/10.1016/0370-1573(88)90023-3}{\emph{Phys. Rept.}
  {\bfseries 168} (1988) 115--207}.

\bibitem{Hu:1993qa}
B.~L. Hu and A.~Matacz, \emph{{Quantum Brownian motion in a bath of parametric
  oscillators: A Model for system - field interactions}},
  \href{https://doi.org/10.1103/PhysRevD.49.6612}{\emph{Phys. Rev.} {\bfseries
  D49} (1994) 6612--6635},
  [\href{https://arxiv.org/abs/gr-qc/9312035}{{\ttfamily gr-qc/9312035}}].

\bibitem{Hu:1986jj}
B.~L. Hu and H.~E. Kandrup, \emph{{Entropy Generation in Cosmological Particle
  Creation and Interactions: A Statistical Subdynamics Analysis}},
  \href{https://doi.org/10.1103/PhysRevD.35.1776}{\emph{Phys. Rev.} {\bfseries
  D35} (1987) 1776}.

\bibitem{Hsiang:2005pz}
J.-T. Hsiang and D.-S. Lee, \emph{{Influence on electron coherence from quantum
  electromagnetic fields in the presence of conducting plates}},
  \href{https://doi.org/10.1103/PhysRevD.73.065022}{\emph{Phys. Rev.}
  {\bfseries D73} (2006) 065022},
  [\href{https://arxiv.org/abs/hep-th/0512059}{{\ttfamily hep-th/0512059}}].

\bibitem{Hsiang:2007zb}
J.-T. Hsiang, T.-H. Wu and D.-S. Lee, \emph{{Stochastic Lorentz forces on a
  point charge moving near the conducting plate}},
  \href{https://doi.org/10.1103/PhysRevD.77.105021}{\emph{Phys. Rev.}
  {\bfseries D77} (2008) 105021},
  [\href{https://arxiv.org/abs/0706.3075}{{\ttfamily 0706.3075}}].

\bibitem{wu3}
T.-H. Wu, J.-T. Hsiang and D.-S. Lee, \emph{Subvacuum effects of the quantum
  field on the dynamics of a test particle},
  {\emph{Annals
  of Physics} {\bfseries 327} (2012) 522 -- 541}.

\bibitem{adscft1}
J.~M. Maldacena, \emph{The large n limit of superconformal field theories and
  supergravity}, {\emph{Adv. Theor. Math. Phys.} {\bfseries 2} (1998)
  231--252}, [\href{https://arxiv.org/abs/hep-th/9711200}{{\ttfamily
  hep-th/9711200}}].

\bibitem{adscft2}
E.~Witten, \emph{Anti-de sitter space and holography}, {\emph{Adv. Theor. Math.
  Phys.} {\bfseries 2} (1998) 253--291},
  [\href{https://arxiv.org/abs/hep-th/9802150}{{\ttfamily hep-th/9802150}}].

\bibitem{Herzog:2006gh}
C.~Herzog, A.~Karch, P.~Kovtun, C.~Kozcaz and L.~Yaffe, \emph{{Energy loss of a
  heavy quark moving through N=4 supersymmetric Yang-Mills plasma}},
  \href{https://doi.org/10.1088/1126-6708/2006/07/013}{\emph{JHEP} {\bfseries
  0607} (2006) 013}, [\href{https://arxiv.org/abs/hep-th/0605158}{{\ttfamily
  hep-th/0605158}}].

\bibitem{Gubser:2006bz}
S.~S. Gubser, \emph{{Drag force in AdS/CFT}},
  \href{https://doi.org/10.1103/PhysRevD.74.126005}{\emph{Phys.Rev.} {\bfseries
  D74} (2006) 126005}, [\href{https://arxiv.org/abs/hep-th/0605182}{{\ttfamily
  hep-th/0605182}}].

\bibitem{CasalderreySolana:2006rq}
J.~Casalderrey-Solana and D.~Teaney, \emph{{Heavy quark diffusion in strongly
  coupled N=4 Yang-Mills}},
  \href{https://doi.org/10.1103/PhysRevD.74.085012}{\emph{Phys.Rev.} {\bfseries
  D74} (2006) 085012}, [\href{https://arxiv.org/abs/hep-ph/0605199}{{\ttfamily
  hep-ph/0605199}}].

\bibitem{Gubser:2006nz}
S.~S. Gubser, \emph{{Momentum fluctuations of heavy quarks in the gauge-string
  duality}},
  \href{https://doi.org/10.1016/j.nuclphysb.2007.09.017}{\emph{Nucl.Phys.}
  {\bfseries B790} (2008) 175--199},
  [\href{https://arxiv.org/abs/hep-th/0612143}{{\ttfamily hep-th/0612143}}].

\bibitem{CasalderreySolana:2007qw}
J.~Casalderrey-Solana and D.~Teaney, \emph{{Transverse Momentum Broadening of a
  Fast Quark in a N=4 Yang Mills Plasma}},
  \href{https://doi.org/10.1088/1126-6708/2007/04/039}{\emph{JHEP} {\bfseries
  0704} (2007) 039}, [\href{https://arxiv.org/abs/hep-th/0701123}{{\ttfamily
  hep-th/0701123}}].

\bibitem{CasalderreySolana:2011us}
J.~Casalderrey-Solana, H.~Liu, D.~Mateos, K.~Rajagopal and U.~A. Wiedemann,
  \emph{{Gauge/String Duality, Hot QCD and Heavy Ion Collisions}},
  \href{https://arxiv.org/abs/1101.0618}{{\ttfamily 1101.0618}}.

\bibitem{Giataganas:2013hwa}
D.~Giataganas and H.~Soltanpanahi, \emph{{Universal Properties of the Langevin
  Diffusion Coefficients}},
  \href{https://doi.org/10.1103/PhysRevD.89.026011}{\emph{Phys.Rev.} {\bfseries
  D89} (2014) 026011}, [\href{https://arxiv.org/abs/1310.6725}{{\ttfamily
  1310.6725}}].

\bibitem{Gursoy:2010aa}
U.~Gursoy, E.~Kiritsis, L.~Mazzanti and F.~Nitti, \emph{{Langevin diffusion of
  heavy quarks in non-conformal holographic backgrounds}},
  \href{https://doi.org/10.1007/JHEP12(2010)088}{\emph{JHEP} {\bfseries 1012}
  (2010) 088}, [\href{https://arxiv.org/abs/1006.3261}{{\ttfamily 1006.3261}}].

\bibitem{Giataganas:2013zaa}
D.~Giataganas and H.~Soltanpanahi, \emph{{Heavy Quark Diffusion in Strongly
  Coupled Anisotropic Plasmas}},
  \href{https://doi.org/10.1007/JHEP06(2014)047}{\emph{JHEP} {\bfseries 06}
  (2014) 047}, [\href{https://arxiv.org/abs/1312.7474}{{\ttfamily 1312.7474}}].

\bibitem{Kovtun:2004de}
P.~Kovtun, D.~T. Son and A.~O. Starinets, \emph{{Viscosity in strongly
  interacting quantum field theories from black hole physics}},
  \href{https://doi.org/10.1103/PhysRevLett.94.111601}{\emph{Phys. Rev. Lett.}
  {\bfseries 94} (2005) 111601},
  [\href{https://arxiv.org/abs/hep-th/0405231}{{\ttfamily hep-th/0405231}}].

\bibitem{Rebhan:2011vd}
A.~Rebhan and D.~Steineder, \emph{{Violation of the Holographic Viscosity Bound
  in a Strongly Coupled Anisotropic Plasma}},
  \href{https://doi.org/10.1103/PhysRevLett.108.021601}{\emph{Phys. Rev. Lett.}
  {\bfseries 108} (2012) 021601},
  [\href{https://arxiv.org/abs/1110.6825}{{\ttfamily 1110.6825}}].

\bibitem{Jain:2015txa}
S.~Jain, R.~Samanta and S.~P. Trivedi, \emph{{The Shear Viscosity in
  Anisotropic Phases}},
  \href{https://doi.org/10.1007/JHEP10(2015)028}{\emph{JHEP} {\bfseries 10}
  (2015) 028}, [\href{https://arxiv.org/abs/1506.01899}{{\ttfamily
  1506.01899}}].

\bibitem{Giataganas:2017koz}
D.~Giataganas, U.~Gürsoy and J.~F. Pedraza, \emph{{Strongly-coupled
  anisotropic gauge theories and holography}},
  \href{https://arxiv.org/abs/1708.05691}{{\ttfamily 1708.05691}}.

\bibitem{deBoer:2008gu}
J.~de~Boer, V.~E. Hubeny, M.~Rangamani and M.~Shigemori, \emph{{Brownian motion
  in AdS/CFT}},
  \href{https://doi.org/10.1088/1126-6708/2009/07/094}{\emph{JHEP} {\bfseries
  0907} (2009) 094}, [\href{https://arxiv.org/abs/0812.5112}{{\ttfamily
  0812.5112}}].

\bibitem{Son:2009vu}
D.~T. Son and D.~Teaney, \emph{{Thermal Noise and Stochastic Strings in
  AdS/CFT}}, \href{https://doi.org/10.1088/1126-6708/2009/07/021}{\emph{JHEP}
  {\bfseries 07} (2009) 021},
  [\href{https://arxiv.org/abs/0901.2338}{{\ttfamily 0901.2338}}].

\bibitem{Lawrence:1993sg}
A.~E. Lawrence and E.~J. Martinec, \emph{{Black hole evaporation along
  macroscopic strings}},
  \href{https://doi.org/10.1103/PhysRevD.50.2680}{\emph{Phys. Rev.} {\bfseries
  D50} (1994) 2680--2691},
  [\href{https://arxiv.org/abs/hep-th/9312127}{{\ttfamily hep-th/9312127}}].

\bibitem{Hawking:1974sw}
S.~W. Hawking, \emph{{Particle Creation by Black Holes}},
  \href{https://doi.org/10.1007/BF02345020}{\emph{Commun. Math. Phys.}
  {\bfseries 43} (1975) 199--220}.

\bibitem{Gibbons:1976pt}
G.~W. Gibbons and M.~J. Perry, \emph{{Black Holes and Thermal Green's
  Functions}}, \href{https://doi.org/10.1098/rspa.1978.0022}{\emph{Proc. Roy.
  Soc. Lond.} {\bfseries A358} (1978) 467--494}.

\bibitem{Unruh:1976db}
W.~G. Unruh, \emph{{Notes on black hole evaporation}},
  \href{https://doi.org/10.1103/PhysRevD.14.870}{\emph{Phys. Rev.} {\bfseries
  D14} (1976) 870}.

\bibitem{Israel:1976ur}
W.~Israel, \emph{{Thermo field dynamics of black holes}},
  \href{https://doi.org/10.1016/0375-9601(76)90178-X}{\emph{Phys. Lett.}
  {\bfseries A57} (1976) 107--110}.

\bibitem{Tong:2012nf}
D.~Tong and K.~Wong, \emph{{Fluctuation and Dissipation at a Quantum Critical
  Point}}, \href{https://doi.org/10.1103/PhysRevLett.110.061602}{\emph{Phys.
  Rev. Lett.} {\bfseries 110} (2013) 061602},
  [\href{https://arxiv.org/abs/1210.1580}{{\ttfamily 1210.1580}}].

\bibitem{Edalati:2012tc}
M.~Edalati, J.~F. Pedraza and W.~Tangarife~Garcia, \emph{{Quantum Fluctuations
  in Holographic Theories with Hyperscaling Violation}},
  \href{https://doi.org/10.1103/PhysRevD.87.046001}{\emph{Phys. Rev.}
  {\bfseries D87} (2013) 046001},
  [\href{https://arxiv.org/abs/1210.6993}{{\ttfamily 1210.6993}}].

\bibitem{Hartnoll:2009ns}
S.~A. Hartnoll, J.~Polchinski, E.~Silverstein and D.~Tong, \emph{{Towards
  strange metallic holography}},
  \href{https://doi.org/10.1007/JHEP04(2010)120}{\emph{JHEP} {\bfseries 04}
  (2010) 120}, [\href{https://arxiv.org/abs/0912.1061}{{\ttfamily 0912.1061}}].

\bibitem{Kiritsis:2012ta}
E.~Kiritsis, \emph{{Lorentz violation, Gravity, Dissipation and Holography}},
  \href{https://doi.org/10.1007/JHEP01(2013)030}{\emph{JHEP} {\bfseries 01}
  (2013) 030}, [\href{https://arxiv.org/abs/1207.2325}{{\ttfamily 1207.2325}}].

\bibitem{Fadafan:2009an}
K.~B. Fadafan, \emph{{Drag force in asymptotically Lifshitz spacetimes}},
  \href{https://arxiv.org/abs/0912.4873}{{\ttfamily 0912.4873}}.

\bibitem{Rajagopal:2015roa}
K.~Rajagopal and A.~V. Sadofyev, \emph{{Chiral drag force}},
  \href{https://doi.org/10.1007/JHEP10(2015)018}{\emph{JHEP} {\bfseries 10}
  (2015) 018}, [\href{https://arxiv.org/abs/1505.07379}{{\ttfamily
  1505.07379}}].

\bibitem{Fischler:2012ff}
W.~Fischler, J.~F. Pedraza and W.~Tangarife~Garcia, \emph{{Holographic Brownian
  Motion in Magnetic Environments}},
  \href{https://doi.org/10.1007/JHEP12(2012)002}{\emph{JHEP} {\bfseries 12}
  (2012) 002}, [\href{https://arxiv.org/abs/1209.1044}{{\ttfamily 1209.1044}}].

\bibitem{Moerman:2016wpv}
R.~W. Moerman and W.~A. Horowitz, \emph{{A semi-classical recipe for wobbly
  limp noodles in partonic soup}},
  \href{https://arxiv.org/abs/1605.09285}{{\ttfamily 1605.09285}}.

\bibitem{Dudal:2018rki}
D.~Dudal and T.~G. Mertens, \emph{{A holographic estimate of heavy quark
  diffusion in a magnetic field}},
  \href{https://arxiv.org/abs/1802.02805}{{\ttfamily 1802.02805}}.

\bibitem{Yeh:2013mca}
C.-P. Yeh, J.-T. Hsiang and D.-S. Lee, \emph{{Holographic Approach to
  Nonequilibrium Dynamics of Moving Mirrors Coupled to Quantum Critical
  Theories}}, \href{https://doi.org/10.1103/PhysRevD.89.066007}{\emph{Phys.
  Rev.} {\bfseries D89} (2014) 066007},
  [\href{https://arxiv.org/abs/1310.8416}{{\ttfamily 1310.8416}}].

\bibitem{Yeh:2015cra}
C.-P. Yeh and D.-S. Lee, \emph{{Subvacuum effects in quantum critical theories
  from a holographic approach}},
  \href{https://doi.org/10.1103/PhysRevD.93.126006}{\emph{Phys. Rev.}
  {\bfseries D93} (2016) 126006},
  [\href{https://arxiv.org/abs/1510.05778}{{\ttfamily 1510.05778}}].

\bibitem{Lee:2017qnr}
D.-S. Lee and C.-P. Yeh, \emph{{Environment-induced uncertainties on moving
  mirrors in quantum critical theories via holography}},
  \href{https://arxiv.org/abs/1706.08283}{{\ttfamily 1706.08283}}.



\bibitem{Roychowdhury:2015mta}
  D.~Roychowdhury,
  ``Quantum fluctuations and thermal dissipation in higher derivative gravity,''
  Nucl.\ Phys.\ B {\bf 897} (2015) 678
    [\href{https://arxiv.org/abs/1506.04548}{{\ttfamily 1506.04548}}].

\bibitem{Banerjee:2013rca}
  P.~Banerjee and B.~Sathiapalan,
  ``Holographic Brownian Motion in 1+1 Dimensions,''
  Nucl.\ Phys.\ B {\bf 884} (2014) 74
  [\href{https://arxiv.org/abs/1308.3352}{{\ttfamily 1308.3352}}].


\bibitem{Motl:2003cd}
L.~Motl and A.~Neitzke, \emph{{Asymptotic black hole quasinormal frequencies}},
  \href{https://doi.org/10.4310/ATMP.2003.v7.n2.a4}{\emph{Adv. Theor. Math.
  Phys.} {\bfseries 7} (2003) 307--330},
  [\href{https://arxiv.org/abs/hep-th/0301173}{{\ttfamily hep-th/0301173}}].

\bibitem{Maldacena:1996ix}
J.~M. Maldacena and A.~Strominger, \emph{{Black hole grey body factors and
  d-brane spectroscopy}},
  \href{https://doi.org/10.1103/PhysRevD.55.861}{\emph{Phys. Rev.} {\bfseries
  D55} (1997) 861--870},
  [\href{https://arxiv.org/abs/hep-th/9609026}{{\ttfamily hep-th/9609026}}].

\bibitem{Harmark:2007jy}
T.~Harmark, J.~Natario and R.~Schiappa, \emph{{Greybody Factors for
  d-Dimensional Black Holes}},
  \href{https://doi.org/10.4310/ATMP.2010.v14.n3.a1}{\emph{Adv. Theor. Math.
  Phys.} {\bfseries 14} (2010) 727--794},
  [\href{https://arxiv.org/abs/0708.0017}{{\ttfamily 0708.0017}}].

\bibitem{Gour:1998my}
G.~Gour and L.~Sriramkumar, \emph{{Will small particles exhibit Brownian motion
  in the quantum vacuum?}},
  \href{https://doi.org/10.1023/A:1018846501958}{\emph{Found. Phys.} {\bfseries
  29} (1999) 1917--1949},
  [\href{https://arxiv.org/abs/quant-ph/9808032}{{\ttfamily
  quant-ph/9808032}}].

\bibitem{Wu:2005jr}
C.-H. Wu and D.-S. Lee, \emph{{Nonequilibrium dynamics of moving mirrors in
  quantum fields: Influence functional and Langevin equation}},
  \href{https://doi.org/10.1103/PhysRevD.71.125005}{\emph{Phys. Rev.}
  {\bfseries D71} (2005) 125005},
  [\href{https://arxiv.org/abs/quant-ph/0501127}{{\ttfamily
  quant-ph/0501127}}].

\bibitem{Dodonov:2010zza}
V.~V. Dodonov, \emph{{Current status of the dynamical Casimir effect}},
  \href{https://doi.org/10.1088/0031-8949/82/03/038105}{\emph{Phys. Scripta}
  {\bfseries 82} (2010) 038105}.

\bibitem{Milton:2004ya}
K.~A. Milton, \emph{{The Casimir effect: Recent controversies and progress}},
  \href{https://doi.org/10.1088/0305-4470/37/38/R01}{\emph{J. Phys.} {\bfseries
  A37} (2004) R209}, [\href{https://arxiv.org/abs/hep-th/0406024}{{\ttfamily
  hep-th/0406024}}].

\bibitem{Giataganas:2015ksa}
D.~Giataganas, \emph{{$k$-Strings as Fundamental Strings}},
  \href{https://doi.org/10.1007/JHEP05(2015)134}{\emph{JHEP} {\bfseries 05}
  (2015) 134}, [\href{https://arxiv.org/abs/1503.09180}{{\ttfamily
  1503.09180}}].

\bibitem{Gubser:2000mm}
S.~S. Gubser and I.~Mitra, \emph{{The Evolution of unstable black holes in
  anti-de Sitter space}},
  \href{https://doi.org/10.1088/1126-6708/2001/08/018}{\emph{JHEP} {\bfseries
  08} (2001) 018}, [\href{https://arxiv.org/abs/hep-th/0011127}{{\ttfamily
  hep-th/0011127}}].

\bibitem{Dong:2012se}
X.~Dong, S.~Harrison, S.~Kachru, G.~Torroba and H.~Wang, \emph{{Aspects of
  holography for theories with hyperscaling violation}},
  \href{https://doi.org/10.1007/JHEP06(2012)041}{\emph{JHEP} {\bfseries 1206}
  (2012) 041}, [\href{https://arxiv.org/abs/1201.1905}{{\ttfamily 1201.1905}}].

\bibitem{Kachru:2008yh}
S.~Kachru, X.~Liu and M.~Mulligan, \emph{{Gravity Duals of Lifshitz-like Fixed
  Points}}, \href{https://doi.org/10.1103/PhysRevD.78.106005}{\emph{Phys.Rev.}
  {\bfseries D78} (2008) 106005},
  [\href{https://arxiv.org/abs/0808.1725}{{\ttfamily 0808.1725}}].

\bibitem{Narayan:2012hk}
K.~Narayan, \emph{{On Lifshitz scaling and hyperscaling violation in string
  theory}}, \href{https://doi.org/10.1103/PhysRevD.85.106006}{\emph{Phys. Rev.}
  {\bfseries D85} (2012) 106006},
  [\href{https://arxiv.org/abs/1202.5935}{{\ttfamily 1202.5935}}].

\bibitem{Azeyanagi:2009pr}
T.~Azeyanagi, W.~Li and T.~Takayanagi, \emph{{On String Theory Duals of
  Lifshitz-like Fixed Points}},
  \href{https://doi.org/10.1088/1126-6708/2009/06/084}{\emph{JHEP} {\bfseries
  0906} (2009) 084}, [\href{https://arxiv.org/abs/0905.0688}{{\ttfamily
  0905.0688}}].

\bibitem{Mateos:2011ix}
D.~Mateos and D.~Trancanelli, \emph{{The anisotropic N=4 super Yang-Mills
  plasma and its instabilities}},
  \href{https://doi.org/10.1103/PhysRevLett.107.101601}{\emph{Phys.Rev.Lett.}
  {\bfseries 107} (2011) 101601},
  [\href{https://arxiv.org/abs/1105.3472}{{\ttfamily 1105.3472}}].


\bibitem{Mateos:2011tv}
D.~Mateos and D.~Trancanelli, \emph{{Thermodynamics and Instabilities of a
  Strongly Coupled Anisotropic Plasma}},
  \href{https://doi.org/10.1007/JHEP07(2011)054}{\emph{JHEP} {\bfseries 1107}
  (2011) 054}, [\href{https://arxiv.org/abs/1106.1637}{{\ttfamily 1106.1637}}].

\bibitem{Jain:2014vka}
S.~Jain, N.~Kundu, K.~Sen, A.~Sinha and S.~P. Trivedi, \emph{{A Strongly
  Coupled Anisotropic Fluid From Dilaton Driven Holography}},
  \href{https://doi.org/10.1007/JHEP01(2015)005}{\emph{JHEP} {\bfseries 01}
  (2015) 005}, [\href{https://arxiv.org/abs/1406.4874}{{\ttfamily 1406.4874}}].

\bibitem{Donos:2016zpf}
A.~Donos, J.~P. Gauntlett and O.~Sosa-Rodriguez, \emph{{Anisotropic plasmas
  from axion and dilaton deformations}},
  \href{https://doi.org/10.1007/JHEP11(2016)002}{\emph{JHEP} {\bfseries 11}
  (2016) 002}, [\href{https://arxiv.org/abs/1608.02970}{{\ttfamily
  1608.02970}}].

\bibitem{Birrelldavies}
N.~D. Birrell and P.~C.~W. Davies, \emph{{Quantum Fields in Curved Space}},
  {\emph{Cambridge Univ. Press.} (1982) 106008}.

\bibitem{Giataganas:2014hma}
D.~Giataganas and K.~Sfetsos, \emph{{Non-integrability in non-relativistic
  theories}}, \href{https://doi.org/10.1007/JHEP06(2014)018}{\emph{JHEP}
  {\bfseries 06} (2014) 018},
  [\href{https://arxiv.org/abs/1403.2703}{{\ttfamily 1403.2703}}].

\bibitem{CaronHuot:2011dr}
S.~Caron-Huot, P.~M. Chesler and D.~Teaney, \emph{{Fluctuation, dissipation,
  and thermalization in non-equilibrium AdS5 black hole geometries}},
  \href{https://doi.org/10.1103/PhysRevD.84.026012}{\emph{Phys. Rev.}
  {\bfseries D84} (2011) 026012},
  [\href{https://arxiv.org/abs/1102.1073}{{\ttfamily 1102.1073}}].

\bibitem{Sonner:2012if}
J.~Sonner and A.~G. Green, \emph{{Hawking Radiation and Non-equilibrium Quantum
  Critical Current Noise}},
  \href{https://doi.org/10.1103/PhysRevLett.109.091601}{\emph{Phys. Rev. Lett.}
  {\bfseries 109} (2012) 091601},
  [\href{https://arxiv.org/abs/1203.4908}{{\ttfamily 1203.4908}}].

\bibitem{Giataganas:2012zy}
D.~Giataganas, \emph{{Probing strongly coupled anisotropic plasma}},
  \href{https://doi.org/10.1007/JHEP07(2012)031}{\emph{JHEP} {\bfseries 1207}
  (2012) 031}, [\href{https://arxiv.org/abs/1202.4436}{{\ttfamily 1202.4436}}].

\bibitem{Chernicoff:2012iq}
M.~Chernicoff, D.~Fernandez, D.~Mateos and D.~Trancanelli, \emph{{Drag force in
  a strongly coupled anisotropic plasma}},
  \href{https://doi.org/10.1007/JHEP08(2012)100}{\emph{JHEP} {\bfseries 1208}
  (2012) 100}, [\href{https://arxiv.org/abs/1202.3696}{{\ttfamily 1202.3696}}].

\bibitem{Giataganas:2013lga}
D.~Giataganas, \emph{{Observables in Strongly Coupled Anisotropic Theories}},
  {\emph{PoS} {\bfseries Corfu2012} (2013) 122},
  [\href{https://arxiv.org/abs/1306.1404}{{\ttfamily 1306.1404}}].

\bibitem{twostate1}
A.~J. Leggett, S.~Chakravarty, A.~T. Dorsey, M.~P.~A. Fisher, A.~Garg and
  W.~Zwerger, \emph{Dynamics of the dissipative two-state system},
  \href{https://doi.org/10.1103/RevModPhys.59.1}{\emph{Rev. Mod. Phys.}
  {\bfseries 59} (Jan, 1987) 1--85}.

\bibitem{twostate2}
R.~Bulla, N.-H. Tong and M.~Vojta, \emph{Numerical renormalization group for
  bosonic systems and application to the sub-ohmic spin-boson model},
  \href{https://doi.org/10.1103/PhysRevLett.91.170601}{\emph{Phys. Rev. Lett.}
  {\bfseries 91} (Oct, 2003) 170601}.

\bibitem{qpt22}
L.~F. Cugliandolo, D.~R. Grempel, G.~Lozano, H.~Lozza and C.~A.
  da~Silva~Santos, \emph{Dissipative effects on quantum glassy systems},
  \href{https://doi.org/10.1103/PhysRevB.66.014444}{\emph{Phys. Rev. B}
  {\bfseries 66} (Jul, 2002) 014444}.

\bibitem{qpt11}
M.~Al-Ali and T.~Vojta, \emph{Quantum phase transition of the sub-ohmic rotor
  model}, \href{https://doi.org/10.1103/PhysRevB.84.195136}{\emph{Phys. Rev. B}
  {\bfseries 84} (Nov, 2011) 195136}.

\bibitem{HoyosBadajoz:2009pv}
C.~Hoyos-Badajoz, \emph{{Drag and jet quenching of heavy quarks in a strongly
  coupled N=2* plasma}},
  \href{https://doi.org/10.1088/1126-6708/2009/09/068}{\emph{JHEP} {\bfseries
  0909} (2009) 068}, [\href{https://arxiv.org/abs/0907.5036}{{\ttfamily
  0907.5036}}].

\end{thebibliography}
\providecommand{\href}[2]{#2}\begingroup\raggedright\endgroup

\end{document}